\documentclass[floatfix,twocolumn,showpacs,floats,aps,pre,showpacs,amsfonts,amssymb]{revtex4-1}
\usepackage{amsmath}
\usepackage{url}
\usepackage{color}
\usepackage{hyphenat}
\usepackage{graphicx}
\usepackage{dcolumn}
\usepackage{bm}
\usepackage{hyperref}

\def\eqref#1{(\ref{#1})}

\newcolumntype{t}[1]{D{.}{.}{#1}}
\newcolumntype{.}{D{.}{.}{-1}}

\begin{document}

\title{Connected component identification and cluster update on GPU}

\author{Martin Weigel}
\affiliation{Institut f\"ur Physik, Johannes Gutenberg-Universit\"at Mainz,
  Staudinger Weg 7, D-55099 Mainz, Germany}
\email[Email me at:] {weigel@uni-mainz.de}
\homepage[Visit: ]{http://www.cond-mat.physik.uni-mainz.de/~weigel}

\date{\today}

\begin{abstract}
  Cluster identification tasks occur in a multitude of contexts in physics and
  engineering such as, for instance, cluster algorithms for simulating spin models,
  percolation simulations, segmentation problems in image processing, or network
  analysis. While it has been shown that graphics processing units (GPUs) can result
  in speedups of two to three orders of magnitude as compared to serial codes on CPUs
  for the case of local and thus naturally parallelized problems such as single-spin
  flip update simulations of spin models, the situation is considerably more
  complicated for the non-local problem of cluster or connected component
  identification. I discuss the suitability of different approaches of
  parallelization of cluster labeling and cluster update algorithms for calculations
  on GPU and compare to the performance of serial implementations.
\end{abstract}


\maketitle

\section{Introduction}

Due to their manifold applications in statistical and condensed matter physics
ranging from the description of magnetic systems over models for the gas-liquid
transition to biological problems, classical spin models have been very widely
studied in the past decades. Since exact solutions are only available for a few
exceptional cases \cite{baxter:book}, with the steady increase in available computer
power and the advancement of simulational techniques, in many cases computer
simulations have become the tool of choice even above the more traditional
variational and perturbative techniques \cite{binder:book2}. The workhorse of Monte
Carlo simulations in the form of the Metropolis algorithm \cite{metropolis:53a} is
extremely general and robust, but suffers from problems of slowed down dynamics in
the vicinity of phase transitions, or for systems with complex free-energy
landscapes. For the case of continuous phase transitions, critical slowing down is
observed with autocorrelation times increasing as $\tau \sim L^z$ with $z\approx 2$
in the vicinity of the critical point. This divergence of temporal correlations is a
consequence of the divergent critical correlations in space, compared to which local
modifications of the configuration become inefficient. An exceptionally successful
solution of this problem is given by a class of cluster-update algorithms working on
stochastically defined connected regions of spins with identical or similar
orientation \cite{swendsen-wang:87a,wolff:89a,kandel:91a,kawashima:95a}, which allow
for a significant reduction of the dynamical critical exponent $z$ over the local
value $z\approx 2$ and can thus easily lead to an effective speed gain in excess of
$10^6$ for practically considered system sizes. Incidentally, the practical task of
cluster identification resulting from the probabilistic description of the problem as
a bond-correlated percolation model is identical to that encountered in image
segmentation or computer vision, where neighboring pixels should be lumped together
according to their colors, a problem which can be mapped to the Potts model
\cite{opara:98,wang:08}. Also, numerical simulations of percolation problems, with
their wide range of realizations from fluids in porous media to epidemic spreading
\cite{stauffer:book}, must deal with a very similar problem of cluster identification
(see, e.g., Ref.~\cite{hoshen:79}). Further applications occur in network analysis,
particle tracking or the identification of structures such as droplets in condensed
matter. Efficient implementations of cluster labeling algorithms are, therefore, of
significant interest for a number of different applications in scientific computing.

In parallel to the invention of new simulation algorithms, the need for strong
computing power for tackling hard problems has prompted scientists to always make the
best use of the available computer resources of the time, be it regular PCs, vector
computers or Beowulf clusters. For the case of simulations of spin models, for
instance, a number of special purpose computers has been devised, including machines
for local updates such as JANUS for spin glasses \cite{belleti:09} and variants such
as the ``cluster processor'' using cluster-update algorithms \cite{bloete:99a}. While
these were (and are) highly successful in their specific application fields, their
design and realization is a rather challenging endeavor, costly in terms of monetary
as well as human resources. It is therefore desirable to search for a less
application specific, but still highly performant platform for massively parallel
scientific computing that is less expensive in terms of its acquisition as well as
its power consumption and cooling requirements than traditional cluster computers. An
architecture meeting those standards has become available in recent years with the
advent of general purpose computing on graphics processing units (GPUs)
\cite{owens:08,kirk:10}. With the availability of convenient application programming
interfaces (APIs) for GPU computing, most notably NVIDIA CUDA and OpenCL
\cite{kirk:10}, the programming effort does no longer dramatically exceed that of CPU
based parallel machines. Still, for efficient implementations architectural
peculiarities of these devices, in particular the organization of compute units
(cores) in groups (multiprocessors) with reduced synchronization capabilities between
multiprocessors and the pyramid of memories with latencies, sizes and access scopes
decreasing from base to tip, need to be taken into account. For the case of spin
models, a wide range of simulation algorithms with local updates has been previously
implemented on GPU \cite{preis:09,bernaschi:10,weigel:10a,ferrero:11}, where for the
implementations reported in Refs.~\cite{weigel:10c,weigel:10a,weigel:11} significant
speedups of two to three orders of magnitude as compared to serial CPU codes have
been reported. An efficient parallelization of non-local algorithms and cluster
labeling is significantly more challenging, however, in particular for the case of
cluster updates for spin models close to criticality, where the relevant clusters
undergo a percolation transition and are therefore spanning the whole system
\cite{heermann:90,baillie:91,mino:91,apostolakis:92,barkema:94,bauernfeind:94,flanigan:95,martin-herrero:04}.

The implementations discussed here have been realized \linebreak within the NVIDIA
CUDA \cite{cuda} framework with benchmarks performed on the GTX 480, GTX 580 and
Tesla M2070 GPUs. While some of the details are specific to this setup, the
algorithmic approaches discussed are fairly general and could easily applied to other
GPU devices or realized with different APIs such as OpenCL. For an introduction into
the details of the GPU hardware and the corresponding programming models, the reader
is referred to the available textbooks (see, e.g, Ref.~\cite{kirk:10}) and previous
articles by the present author \cite{weigel:10c,weigel:10a}.

The rest of the paper is organized as follows. In Section II GPU implementations of
the cluster algorithm of Swendsen and Wang \cite{swendsen-wang:87a} are
discussed. The cluster decomposition of the complete spin lattice necessary here is
identical to that of a corresponding image segmentation problem or percolation
simulation. Section III is devoted to the case of the single-cluster variant
suggested by Wolff \cite{wolff:89a}. Finally, Section IV contains my conclusions.

\section{Swendsen-Wang algorithm\label{sec:SW}}

In this paper, I focus on the ferromagnetic \(q\)-state Potts model with Hamiltonian
\begin{equation}
  \label{eq:potts_hamiltonian}
  {\cal H} = -J\sum_{\langle ij\rangle} \delta_{s_i s_j},
\end{equation}
where $s_i \in \{1,\ldots, q\}$ denote the spin variables, $J > 0$ is the exchange
coupling, and the sum extends over all bonds of an underlying graph, most commonly a
regular lattice. In dimensions $d> 1$, the model undergoes a transition from a
disordered phase at high temperatures to an ordered phase where one of the $q$ states
prevails at low temperatures \cite{wu:82a}. For $d=2$, the transition is continuous
for $q\le 4$ and first order for $q> 4$, while in $d=3$ it is first order for any
$q\ge 3$. The special case $q=2$ is equivalent to the celebrated Ising model. A local
Monte Carlo simulation of the Potts model proceeds by iteratively changing the
orientation of randomly chosen spin variables in accordance with the detailed balance
condition \cite{binder:book2}. In contrast, the cluster algorithm of Swendsen and
Wang \cite{swendsen-wang:87a} updates connected components of (usually) more than one
spin and is based on the following transformation of the partition function due to
Fortuin and Kasteleyn \cite{fortuin:72a},
\begin{eqnarray}
  \label{eq:fortuin1}
  Z & = & \sum_{\{s_i\}}\exp\left(\beta J\sum_{\langle ij\rangle} \delta_{s_i s_j}\right) \\
    \label{eq:fortuin2}
    & = & \sum_{\{s_i\}} \prod_{\langle ij\rangle} e^{\beta J} \left[(1-p)+p\delta_{s_i
      s_j}\right]\\
  \label{eq:fortuin3}
    & = &  \sum_{\{n_{ij}\}} \sum_{\{s_i\}} \prod_{\langle ij\rangle} 
    e^{\beta J}\left[(1-p)\delta_{n_{ij,0}}+p\delta_{s_i s_j}\delta_{n_{ij},1}\right], 
\end{eqnarray}
where $\beta$ denotes the inverse temperature and $p = 1-e^{-\beta J}$. In
Eq.~\eqref{eq:fortuin3}, a set of auxiliary bond variables $n_{ij} \in \{0,1\}$ is
introduced, where $n_{ij} = 0$ whenever $s_i \ne s_j$ and $n_{ij} = 1$ with
probability $p$ for $s_i = s_j$. The resulting stochastically defined clusters are
therefore subsets of the geometric clusters of parallel spins. Using a graphical
expansion of the term in square brackets in Eq.~\eqref{eq:fortuin3} and summing over
the spin configurations $\{s_i\}$, it can be shown that the model is equivalent to a
generalized percolation model with partition function \cite{fortuin:72a,hu:84a}
\begin{equation}
  \label{eq:random_cluster}
  Z = e^{\beta J}\sum_{\{n_{ij}\}} p^{b(\{n_{ij}\})}(1-p)^{{\cal
      E}-b(\{n_{ij}\})}q^{n(\{n_{ij}\})},
\end{equation}
known as the random-cluster model. Here, $b(\{n_{ij}\})$ denotes the number of
activated edges resulting from the bond variables $n_{ij}$, $n(\{n_{ij}\}$ is the
number of connected components of the induced subgraph, and ${\cal E}$ is the total
number of edges in the underlying graph or lattice. From the percolation
representation \eqref{eq:random_cluster} it is clear \cite{coniglio:80a} that the
stochastic clusters induced by the bond variables $n_{ij}$ (and not the geometric
clusters of like spins) undergo a percolation transition at the thermal transition
point, and hence it is these structures that should be updated to efficiently
decorrelate the system close to criticality.

Utilizing the representation \eqref{eq:fortuin3} the algorithm by Swendsen and Wang
alternatingly updates spins $s_i$ and bond variables $n_{ij}$ as follows:
\begin{enumerate}
\item For a given spin configuration set $n_{ij} = 0$ for each bond with $s_i \ne
  s_j$. Set $n_{ij} = 1$ and $n_{ij} = 0$ with probabilities $p$ and $1-p$,
  respectively, for each bond with $s_i = s_j$.
\item Identify the connected components of the subgraph of the lattice induced by the
  bond variables $n_{ij}$.
\item Choose a new spin orientation randomly in $\{1,\ldots,q\}$ for each connected
  component and update the spin variables $s_i$ accordingly.
\end{enumerate}
Since clusters of single spins are possible, this update is trivially ergodic. It is
straightforward to show that detailed balance is fulfilled
\cite{swendsen-wang:87a,kawashima:95a}. Hence, the Swendsen-Wang (SW) dynamics forms
a valid Markov chain Monte Carlo algorithm of the Potts model. Autocorrelations are
dramatically reduced as compared to local spin flips. A rigorous bound for the
dynamical critical exponent is $z_{\mathrm{int}} \ge \alpha/\nu$ \cite{li:89}, where
$z_{\mathrm{int}}$ is the exponent of the scaling of the integrated autocorrelation
time and $\alpha$ and $\nu$ are the (static) critical exponents of the specific heat
and the correlation length, respectively. This bound is close to sharp in two
dimensions \cite{deng:07a}, but not in $d=3$ where, nevertheless, significant
reductions in autocorrelations and the dynamical critical exponent $z$ are observed.

Attempting a highly parallel GPU implementation of the SW algorithm, it is clear that
the bond activation in step 1 as well as the cluster flipping in step 3 can be rather
easily parallelized as they are perfectly local operations. In contrast, the cluster
identification in step 3 must deal with structures spanning the whole system, in
particular for simulations close to criticality which are the main strength of
cluster updates. This is also the crucial step for further applications of cluster
identification such as the image segmentation problem mentioned above. The total run
time for a single update of the spin lattice with the SW algorithm on a single GPU
therefore decomposes as
\begin{equation}
  T^p_\mathrm{SW} = T^p_\mathrm{activate} + T^p_\mathrm{identify} + T^p_\mathrm{flip}.
\end{equation}
We distinguish these times from the corresponding serial run times $T^s_\mathrm{SW}$,
$T^s_\mathrm{activate}$, etc.\ for single-threaded calculations. For definiteness,
the implementation is discussed in some detail for the specific example of the Potts
model on the square lattice of edge length $L$ with periodic boundary
conditions. Generalizations to three dimensions or other lattice types are
straightforward.

\subsection{Bond activation}

We use an array of $2L^2$ {\tt char} variables to represent the bond activation
states $n_{ij}$. For the GPU implementation using CUDA \cite{kirk:10}, bond
activation is performed by a first kernel, {\tt prepare\_bonds()}. Given a
configuration of the spins $s_i$, for each bond an expression of the form
\begin{equation}
  n_{ij} = \left\{
    \begin{array}{cl}
      1 & \mbox{if $s_i = s_j$ and $r < p$}\\
      0 & \mbox{otherwise}
    \end{array}
  \right.
\end{equation}
needs to be evaluated, where $r\in(0,1)$ is a uniform (pseudo) random number, and $p
= 1-e^{-\beta J}$. To enable parallelism, the system is broken up into tiles of $B^2
= B_x\times B_y$ spins, and each tile is assigned to an independent thread block. If
we denote $\ell_x = L/B_x$, $\ell_y = L/B_y$ and $\ell^2 = \ell_x\ell_y$ the number
of tiles, the expected parallel run time behaves as
\begin{equation}
  T^p_\mathrm{activate} \sim \frac{\ell^2}{\min(\ell^2, n)} \frac{B^2}{\min(B^2/k,m)},
  \label{eq:activate_scaling}
\end{equation}
where $n$ denotes the number of multiprocessors ($n=14$ for Tesla M2070, $n=15$ for
GTX 480, $n=16$ for GTX 580), $m$ is the number of cores per multiprocessor ($n=32$
for all three cards), and $k$ is the number of sites assigned to each
thread\footnote{I do not take the effects of latency hiding and other scheduling
  specificities into account in the scaling formulae, but discuss them in some places
  in connection with observed deviations from these simplified laws. It is also
  assumed that the number of threads per block is at least $4$ since due to the
  limitation to eight active blocks per multiprocessor on current NVIDIA GPUs, there
  would otherwise be idle cores.}. For large systems, $\ell^2 > n$ and $B^2/k > m$,
Eq.~\eqref{eq:activate_scaling} reduces to $T^p_\mathrm{activate} \sim \ell^2 B^2 =
L^2$. As discussed in detail in Refs.~\cite{weigel:10c,weigel:10a}, each thread
requires its own instance of a random number generator (RNG) to prevent the formation
of a performance bottleneck. Due to the resulting large number of RNG instances (for
the case of large systems), one requires a generator with a small state comprising,
ideally, not more than a few bytes. This precludes the use of high-quality, but large
state generators such as Mersenne twister \cite{matsumoto:98} in applications of the
type considered here. Additionally, one needs to ensure that the thus created streams
of random numbers are sufficiently uncorrelated with each other. Suitable generator
types for this purpose are, for instance, arrays of linear congruential generators
with random seeds, which are fast but might not produce random numbers of sufficient
quality \cite{gentle:03,preis:09,weigel:10a}, generalized lagged Fibonacci generators
\cite{weigel:10a}, or the Marsaglia generator as suggested in
Ref.~\cite{ferrero:11}. As the cluster identification step, which does not require
random numbers, dominates the parallel runtime of the algorithm, RNG speed is
not as important as in local update simulations on GPUs. For the benchmarks reported
below, I used an array of 32-bit linear congruential generators. Statistically
significant deviations from the exact results \cite{ferdinand:69a} for the $q=2$
Potts model at criticality have not been observed.

An analysis of the kernel with CUDA's Compute Visual Profiler \cite{cuda} shows that
its performance is compute bound. Still, memory performance can be improved by using
an appropriate memory layout ensuring that reads of subsequent threads in the spin
and bond arrays map to consecutive locations in global memory to ensure coalescence
of memory requests \cite{kirk:10}. With a linear memory arrangement these
requirements are best met when using tiles with $B_x\gg B_y$. Best results for
systems with $L > 256$ are found here for $B_z = 256$, $B_y = 4$ (considering only
lattice sizes $L=2^n$, $n\in\mathbb{N}$). Evaluating the acceptance criterion leads
to unavoidable thread divergence, but the effects are not very dramatic here. The
asymptotic performance of the kernel with one spin per thread, $k=1$, is
$T^p_\mathrm{activate}/L^2 = 0.66$~ns on the GTX 480 (assuming full loading of the
multiprocessors which is reached for sufficiently large systems). An alleviating
effect on thread divergence and memory limitations is reached by assigning several
spin pairs (bonds) to each thread. Two versions have been considered here, either
assigning a (square) sub-block of four spins to each thread ($k=4$), which brings
down the updating time to $T^p_\mathrm{activate}/L^2 = 0.46$~ns, or assigning only
$B_x$ threads per tile, each of which has to update $k=B_y$ spins, leading to the
same asymptotic performance of $T^p_\mathrm{activate}/L^2 = 0.46$~ns on the GTX
480. The better performance of these variant kernels has to do with the possibility
of pre-fetching of data into registers while arithmetic operations are being
performed. The same kernel is used to also initialize the cluster labels (see
below). Note that the relatively lower performance of this kernel as compared to the
Metropolis update of the Ising model reported in Ref.~\cite{weigel:10a} of about
$0.13$~ns per spin (without multi-hit updates) on the same hardware is explained by
the 6-fold increase in memory writes (two {\tt char}s and one {\tt int} versus one
{\tt char}) and the use of two random numbers (instead of one) per spin.

\subsection{Cluster labeling on tiles}

To allow for an efficient use of parallel hardware, cluster labeling is first
performed on (square) tiles of $B\times B$ spins and cluster labels are consolidated
over the whole lattice in a second step (see Sec.~\ref{sec:consolidate} below)
\cite{heermann:90,baillie:91,mino:91,barkema:94,bauernfeind:94,flanigan:95}. Hence
the time for cluster identification naturally breaks up into two contributions,
\begin{equation}
  T^p_\mathrm{identify} = T^p_\mathrm{local} + T^p_\mathrm{global}.
\end{equation}
In the field of simulations of spin systems (and percolation), the standard technique
for cluster identification is due to Hoshen and Kopelman \cite{hoshen:79}. Although
originally not formulated in this context, it turns out to be a special version of
the class of union-and-find algorithms well known in theoretical computer science
\cite{cormen:09}. Time and storage requirements for this approach scale linearly with
the number $N = L^2$ of sites. A somewhat more ``natural'' approach consists of a set
of breadth-first searches on the graph of bonds, growing the clusters in a layered
fashion. While storage requirements are super-linear in $N$ (and might be as large as
$N^2$ depending on the structure of the underlying graph), computing time scales
still linear in $N$ and implementations are typically very straightforward and
efficient. A third approach considered here, dubbed self-labeling \cite{baillie:91},
is very inefficient regarding (serial) computing time, but very well suited for
parallelization.

\begin{figure}[tb]
  \centering
  \includegraphics[width=0.7\columnwidth]{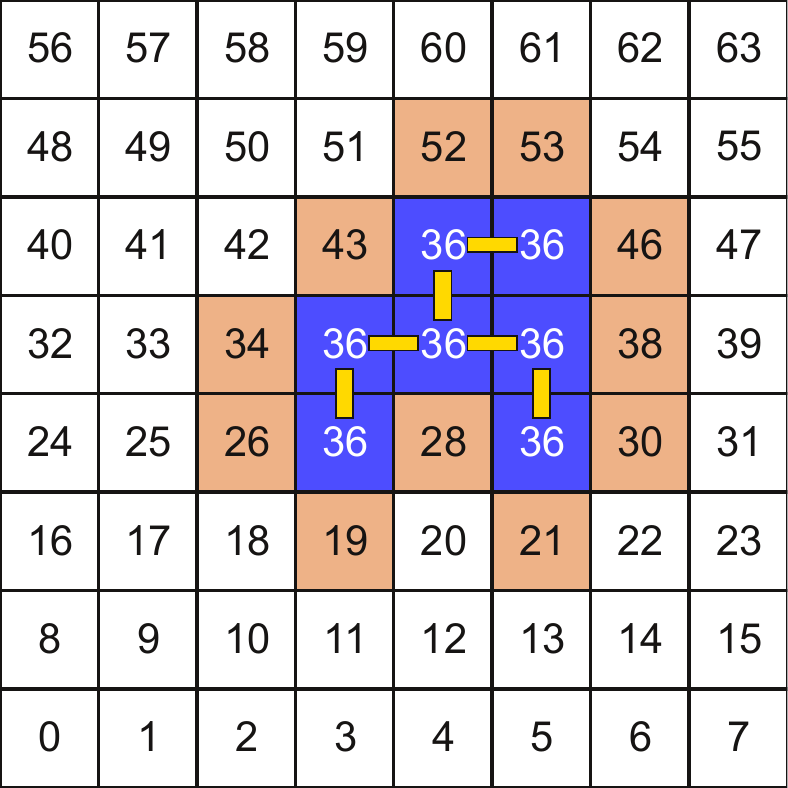}
  \caption{(Color online) Cluster identification on a $64\times 64$ tile using a breadth-first
    search. The already labeled sites are indicated in blue (dark squares) while the
    current wave front of unvisited neighbors is shaded in red (light squares).}
  \label{fig:BFS}
\end{figure}

\subsubsection{Breadth-first search}

In breadth-first search (or ``ants in the labyrinth'') the unvisited neighbors of a
starting vertex or seed that are connected by activated bonds are examined and stored
in a first-in-first-out (FIFO) data structure (a queue). Subsequently, nodes are
removed from the queue in the order they have been stored and examined in the same
fashion as the initial vertex. This leads to a layered growth of the identified part
of a cluster as illustrated in Fig.~\ref{fig:BFS}. The complete set of clusters is
being identified by seeding a new cluster at each node of the lattice that is not yet
part of a previously identified cluster. Information about the cluster structure is
stored in an array of cluster labels, where originally each cluster label is
initialized with the site number on the lattice and cluster labels are set to that of
the seed site on growing the cluster, cf.\ Fig.~\ref{fig:BFS}. While this approach is
very general (it can be applied without changes to any graph) and well suited for
serial calculations, it is not very suitable for parallelization
\cite{bader:06}. Parallelism can be implemented in the examination of different
neighbors of a site and in processing the sites on the wave front of the growing
cluster. To avoid race conditions and achieve consistency of the data structures,
however, locks or atomic operations are required, considerably complicating the
code. Additionally, the number of (quasi) independent tasks is highly variable as the
length of the wave fronts is fluctuating quite strongly. For the case of a parallel
identification of {\em all\/} clusters as necessary for the SW algorithm and image
segmentation, this approach is hence not very well suited for a GPU implementation. A
parallel implementation will be discussed below in the context of the single-cluster
(or Wolff) variant of the algorithm in Sec.~\ref{sec:single}, however.

\begin{figure}[tb]
  \centering
  \includegraphics[keepaspectratio=true,scale=0.8,trim=45 48 75 78]{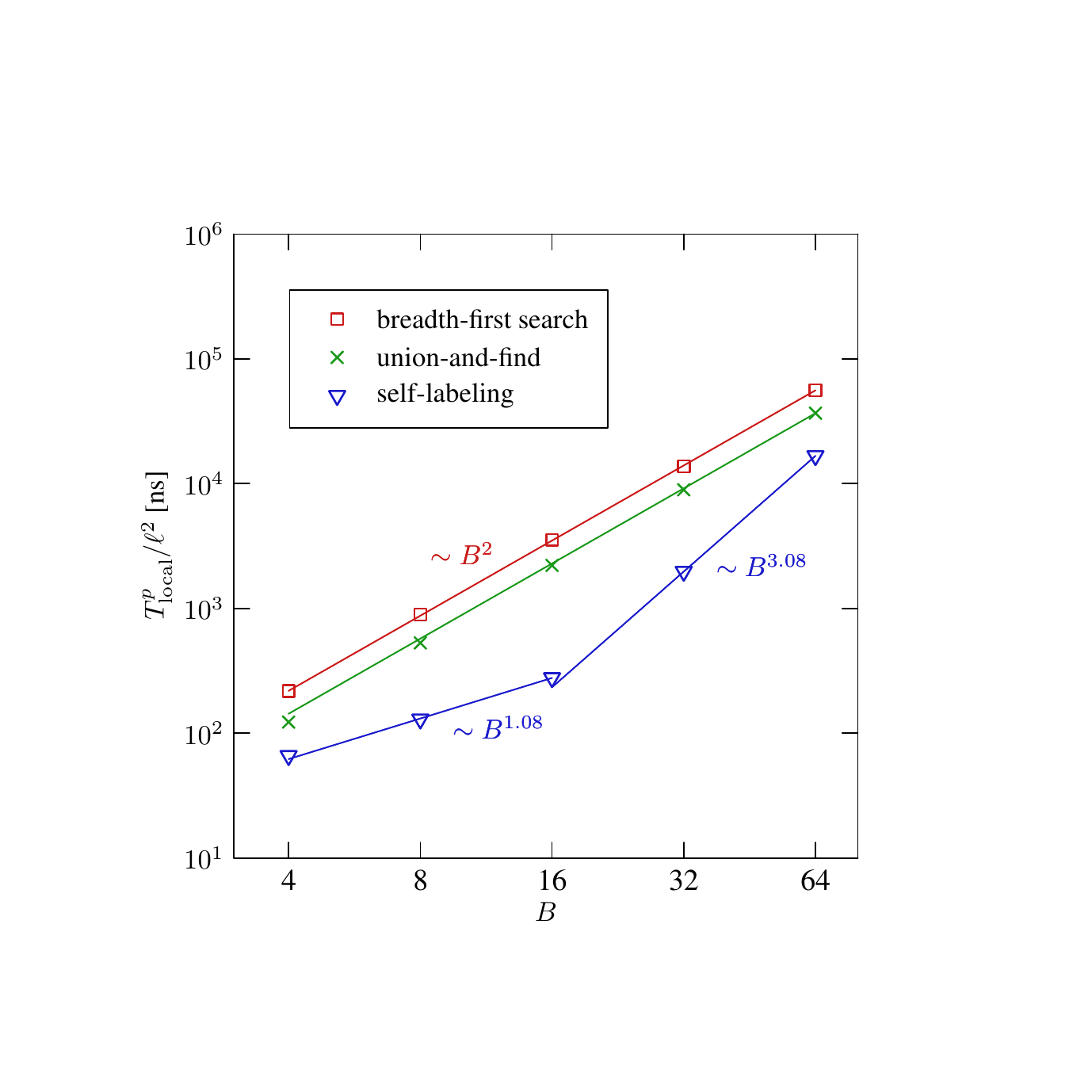}
  \caption{(Color online) Parallel average run time for local cluster labeling on a $4096\times
    4096$ square lattice in tiles of edge length $B$. Data are for the $q=2$ states
    Potts model at the critical point. Breadth-first search and tree-based
    union-and-find are (up to logarithmic corrections) proportional to the number
    $B^2$ of sites, while self-labeling exhibits scaling proportional to
    $B^{2+d_\mathrm{min}} \approx B^{3.08}$. The weaker scaling proportional to
    $B^{d_\mathrm{min}} \approx B^{1.08}$ of self-labeling for small $B$ is due to
    under-utilization of GPU cores (see main text). The lines are fits of the power
    law $T^p_\mathrm{local}/\ell^2 = A B^\kappa$ with the indicated fixed exponents to
    the data.}
  \label{fig:label_scaling}
\end{figure}

The parallel run time of this kernel, {\tt local\_BFS()}, employing one thread per
block performing cluster identification in a tile of edge length $B$, is therefore
expected to scale as
\begin{equation}
  \label{eq:BFS_scaling}
  T^p_\mathrm{local} \sim \frac{\ell^2}{\min(\ell^2, n)} B^2.
\end{equation}
The measured run times for $\ell^2 > n$ follow this expectation, resulting in
perfectly linear scaling of the time $T^p_\mathrm{local}/\ell^2$ per tile with the
number $B^2$ of tile sites, cf.\ Fig.~\ref{fig:label_scaling}. Since only a maximum
of 8 thread blocks can be simultaneously active on each multiprocessor on current
generation NVIDIA GPUs \cite{kirk:10}, however, 24 of the 32 cores of each
multiprocessor are idling, leading to rather low performance. The asymptotic maximum
performance for large system sizes (leading to an optimum effect of latency hiding
through the scheduler) on the GTX 480 is at around $T^p_\mathrm{local}/L^2 = 13.4$~ns
for this kernel, {\tt local\_BFS()}.

\subsubsection{Union-and-find algorithms}

\begin{figure}[tb]
  \centering
  \includegraphics[width=0.7\columnwidth]{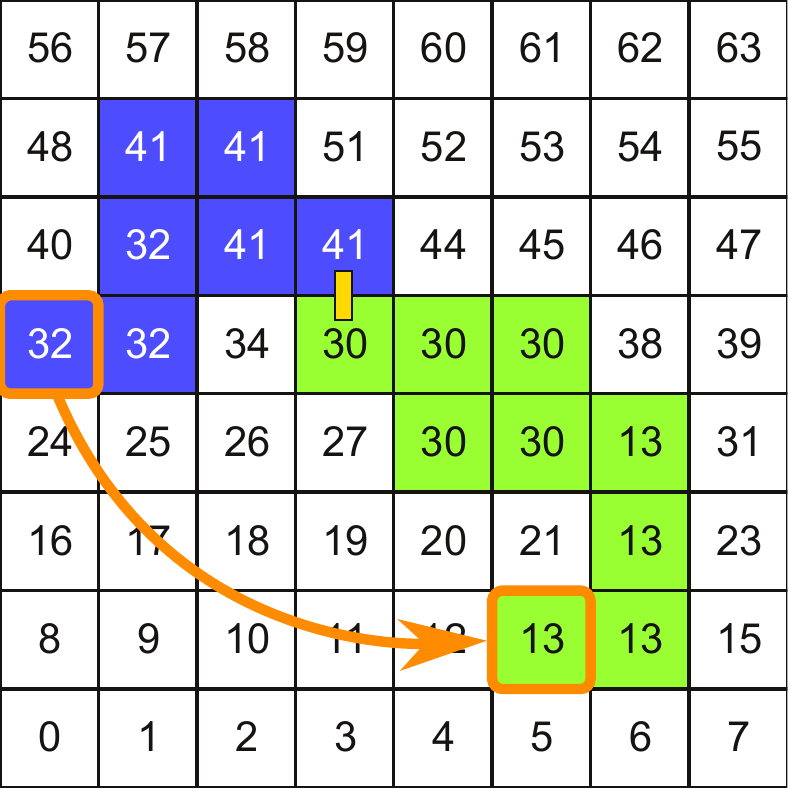}
  \caption{(Color online) Cluster labeling using union-and-find with balanced trees and (partial)
    path compression on a $64\times 64$ tile. Under insertion of the edge between
    sites 30 and 41, the smaller cluster with root No.\ 32 is attached to the root of
    the larger cluster at No.\ 13.}
  \label{fig:unionfind}
\end{figure}

It is a well known problem in computer science to partition a set of elements into
disjoint subsets according to some connectedness or identity criterion. A number of
efficient algorithm for this so-called union-and-find problem have been developed
\cite{cormen:09}. Consider a set of $N$ elements denoted as vertices in a graph
which, initially, has no edges. Now, a number of edges is sequentially inserted into
the graph and the task is to successively update a data structure that contains
information about the connected components resulting from the edge
insertion. Obviously, our cluster identification task is a special case of this
problem. In a straightforward implementation one maintains a forest of spanning trees
where each node carries a pointer to its parent in the tree, unless it is the tree
root which points to itself. On insertion of an edge one determines the roots of the
two adjacent vertices by successively walking up the respective tree structures ({\em
  find\/}). If the two roots found are the same, the inserted edge was internal and
no further action is required. If two different roots were found, the edge was
external and one of the trees is attached to the root of the other as a new branch
({\em union\/}), thus amalgamating two previously disjoint subsets or connected
components of the graph. The forest structure can be realized with an array of node
labels, where each node is initialized to point to itself (i.e., it is its own
root). This process is illustrated for the present application in
Fig.~\ref{fig:unionfind}.

(Worst case) time complexity is trivially constant or $O(1)$ for {\em union\/} steps,
while {\em find\/} steps can be extensive, $O(N)$, if edges connecting macroscopic
clusters are considered. (Storage requirements are clearly just linear in $N$.) The
complexity of the find step can be reduced by two tricks, tree balancing and path
compression. Balancing can be achieved by making sure that always the smaller tree
(in terms of the number of nodes) is attached to the larger. To this end, the current
number of nodes is stored in the tree root. Balancing reduces the time to find the
root to $O(\log N)$ steps \cite{cormen:09}. Similarly path compression, which
redirects the ``up'' pointer of each node to point directly to the tree root in a
backtracking operation after each completed find task, reduces find complexity to
$O(\log N)$. The combination of both techniques can be shown to result in an
essentially constant find complexity for all practically relevant system sizes
\cite{tarjan:75}. An implementation of the full algorithm geared towards cluster
identification is described in Ref.~\cite{newman:01a}.

Like the breadth-first search, the tree-based union-and-find approach is
intrinsically serial as all operations work on the same forest structure, whose
consistency could not be easily maintained under parallel operations. Moderate
parallelsim is possible in the union step, where the two find operations for the
vertices connected by the new edge can be performed in parallel. Due to the resultant
thread divergence, however, using two threads per block is found to actually decrease
performance. Similarly, the extra effect of path compression (keeping the stack for
backtracking in fast shared memory) is found to actually {\em increase\/} run times,
at least in the range of tile sizes $4\le B\le 64$ considered. The parallel scaling
of the algorithm is thus the same (up to logarithmic corrections) as that of
breadth-first search given in Eq.~\eqref{eq:BFS_scaling}. In fact,
$T^p_\mathrm{local}/\ell^2$ is found to be almost perfectly linear in $B^2$ in the
considered regime, cf.\ the data presented in Fig.~\ref{fig:label_scaling}.  The
asymptotic performance (neglecting logarithmic terms due to the find step) of the
kernel {\tt local\_unionfind()} on the GTX 480 is found to be $T^p_\mathrm{local}/L^2
= 8.6$~ns, somewhat better than for {\tt local\_BFS()}. Note that for the tree-based
algorithms of the union-and-find type, memory accesses are inherently non-local
leading to a certain performance penalty which hardly can be avoided.
  
\subsubsection{Self-labeling}

\begin{figure}[tb]
  \centering
  \includegraphics[width=0.7\columnwidth]{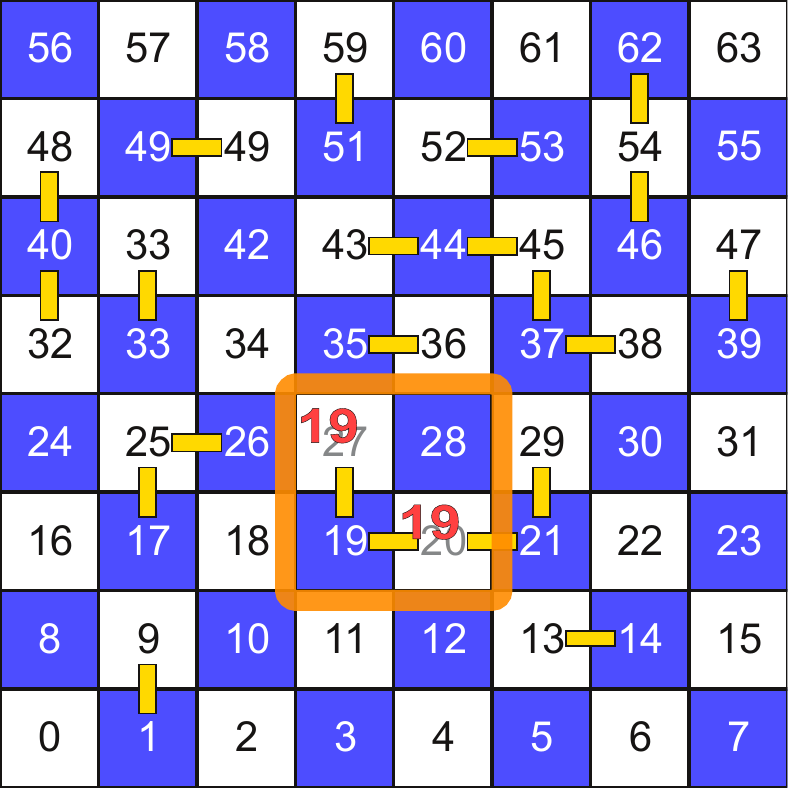}
  \caption{(Color online) Cluster identification on a $64\times 64$ tile using the self-labeling
    algorithm with one thread per $2\times 2$ spins. In every pass, each site examines
    its northward and eastward neighbors and, if they are connected by an active
    bonds, for each pair sets both labels to the minimum of the two current labels.}
  \label{fig:selflabel}
\end{figure}

While breadth-first search and tree-based union-and-find are elegant and very
efficient for serial implementations, they appear not very suitable for
parallelization, especially on GPUs where groups of threads are executed in perfect
synchrony or lockstep and extensive thread divergence is expensive. An antipodal type
of algorithm is given by the simple approach of self-labeling \cite{baillie:91}:
Cluster labels are initialized with the site numbers. Each site then compares its
label to that of its eastward neighbor and sets its own and this neighbor's labels to
the minimum of the two, provided that an activated bond connects the two sites. The
same is subsequently done with respect to the northward neighbor, cf.\
Fig.~\ref{fig:selflabel}. (On a simple cubic lattice, the analog procedure would
involve three out of six neighbors.)  Clearly, the outcome of a whole sweep of
re-labeling events will depend on the order of operations and several passes through
the tile will be necessary until the final cluster labels have propagated through the
whole system. Eventually, however, no label will have changed in a whole pass through
the tile and the procedure can be stopped, leading to a correct labeling of clusters
inside of each tile. Let us first concentrate on the spin model at criticality. Then,
clusters typically span the tile, such that at least of the order of $B$ sweeps will
be required to pass information about correct cluster labels from one end of the tile
to the other. In fact, even more passes are necessary, as information about cluster
labels needs to be diffusively transmitted between each pair of sites in the same
cluster. Since under the chosen dynamics this information will be transmitted along
the shortest path connecting the two sites, the required number of sweeps will scale
as
\begin{equation}
  n_B \sim B^{d_\mathrm{min}},
\end{equation}
where $d_\mathrm{min} \ge 1$ is the fractal dimension of the shortest path on a
percolation cluster. For pure percolation (corresponding to the $q\rightarrow 1$
limit of the Potts model) it is found to be $d_\mathrm{min} \approx 1.13$ in $d=2$
and $d_\mathrm{min} \approx 1.34$ in $d=3$ \cite{herrmann:88}, whereas for the
(Fortuin-Kasteleyn clusters of the) $q=2$ and $q=3$ Potts models in two dimensions it
has been estimated as $d_\mathrm{min} = 1.08(1)$ and $d_\mathrm{min} = 1.01(1)$,
respectively \cite{miranda:91}. Obviously, the approach can be easily parallelized
inside of tiles, assigning an individual thread to one or $k > 1$ spins. As a
consequence, the parallel run time for the self-labeling approach is
\begin{equation}
  T^p_\mathrm{local} = C_\mathrm{local} \frac{\ell^2}{\min(\ell^2, n)} \frac{B^2}{\min(B^2/k,m)}
  B^{d_\mathrm{min}}
  \label{eq:selflabel_scaling}
\end{equation}
at or close to the percolation transition, which asymptotically appears to be rather
unflattering as compared to the breadth-first search and union-and-find
techniques. Due to the parallelization on the tile level, however, the total run time
can still be quite low for intermediate tile sizes. Off criticality, the scaling
becomes somewhat more favorable. Below the transition, where clusters span the
lattice, but they are no longer fractal, $d_\mathrm{min}$ should be replaced by
one. Above the transition, on the other hand, with a finite correlation length $\xi$,
$B^{d_\mathrm{min}}$ in Eq.~\eqref{eq:selflabel_scaling} is replaced by $\min(\xi,
B)$. While this somewhat better behavior is probably not very relevant for the spin
models as simulations close to criticality are the main purpose of cluster-update
algorithms, it is of importance for percolation simulations or image segmentation
problems for the (typical) case of a finite characteristic length scale $\xi$.

Figure \ref{fig:label_scaling} shows the scaling of parallel run times for the kernel
{\tt local\_selflabel()} on tiles of sizes $4\le B \le 64$ for the $q=2$ Potts model
at the critical point $\beta_c = \ln(1+\sqrt{2})$. One can clearly distinguish two
regimes with scaling $T^p_\mathrm{local}/\ell^2 \sim B^{d_\mathrm{min}} \approx
B^{1.08}$ for $B^2/k < m$ and $T^p_\mathrm{local}/\ell^2 \sim B^{2+d_\mathrm{min}}
\approx B^{3.08}$ for $B^2/k > m$. (The data in Fig.~\ref{fig:label_scaling} are for
$k=4$ on the GTX 480 with $m=32$, such that the crossover occurs at $B \approx 11$.)
As is apparent from Fig.~\ref{fig:label_scaling}, for tile sizes $B\le 64$
self-labeling is clearly superior in parallel performance on GPU as compared to
breadth-first search or union-and-find, although it becomes less efficient than the
latter two approaches for $B\gtrsim 128$. I tested different variants: (a) an
implementation, {\tt local\_selflabel\_small()}, that assigns one spin per thread,
restricting the tile size to $B\le 32$ on current NVIDIA GPUs with a limitation of
1024 threads per block, (b) a kernel {\tt local\_selflabel()} which assigns a
$2\times 2$ block of spins to each thread, allowing tile sizes up to $B=64$, and (c)
a looped version, {\tt local\_selflabel\_block()}, that assigns one column of height
$B$ to each thread, such that the lines are worked on through a loop. In all cases,
the relevant portion of the bond activation variables and cluster labels are copied
to shared memory, such that memory fetches in the re-labeling steps are (almost)
instantaneous. Bank conflicts are avoided through an appropriate layout of the data
in shared memory. Depending on the number of spins per thread, a different order of
operations can lead to different results for each single self-labeling
pass. Consistency could be enforced via atomic operations, but these slow down the
code and are found to be not necessary here. Therefore, while the number of necessary
self-labeling passes might vary from run to run (or device to device) depending on
scheduling specificities, the final result is deterministic and does not depend on
the order of operations. The decision about the end of self-labeling is taken using
the warp vote function {\tt \_\_syncthreads\_or()} \cite{cuda} which evaluates to
true as long as {\em any\/} of the threads has seen a re-labeling event in the last
pass. Performance differences between the mentioned three kernels are found to be
relatively small. The best asymptotic performance is observed for the kernel {\tt
  local\_selflabel()} with $2\times 2$ spins per thread, as this setup avoids
read/write conflicts in shared memory. For tiles of size $B=16$ on the GTX 480 the
run time per spin is $T_\mathrm{local}/L^2 = 1.08$~ns for all labeling passes. While
the total number of operations is larger for self-labeling than for breadth-first
search or union-and-find, the former is 13 and 8 times faster than the latter at
$B=16$, respectively, due to the easily exploited inherent parallelism.

\subsection{Tile consolidation\label{sec:consolidate}}

Each of the three cluster labeling algorithms on tiles discussed above results in
correct cluster labels inside of tiles, however, ignoring the information of any
active bonds crossing tile boundaries. To reach unified labels for clusters spanning
several tiles, an additional consolidation phase is necessary. Two alternatives, an
iterative relaxation procedure and a hierarchical sewing scheme have been considered
to this end.

\subsubsection{Label relaxation\label{sec:relax}}

\begin{figure}[tb]
  \centering
  \includegraphics[width=0.7\columnwidth]{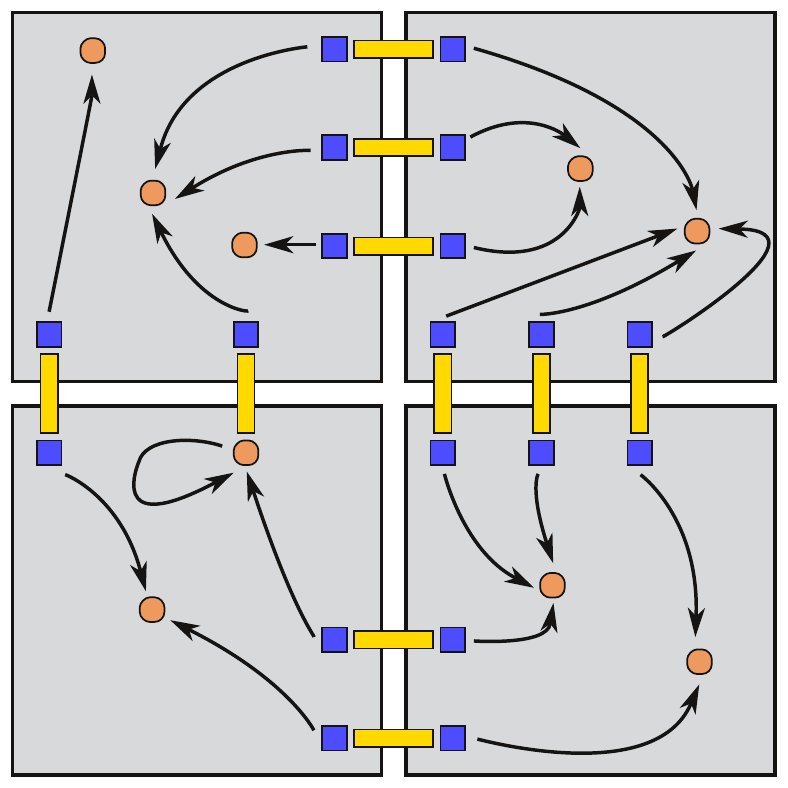}
  \caption{(Color online) Tile consolidation via label relaxation. For each spin on the boundary of
    a tile (squares) with an off-tile active bond, the local root nodes (circles) are
    stored in an array. The corresponding local root labels are transmitted to
    neighboring tiles, who change their local labels to the minimum of their own and
    the received labels.}
  \label{fig:relax}
\end{figure}

\begin{figure}[tb]
  \centering
  \includegraphics[keepaspectratio=true,scale=0.8,trim=45 48 75 78]{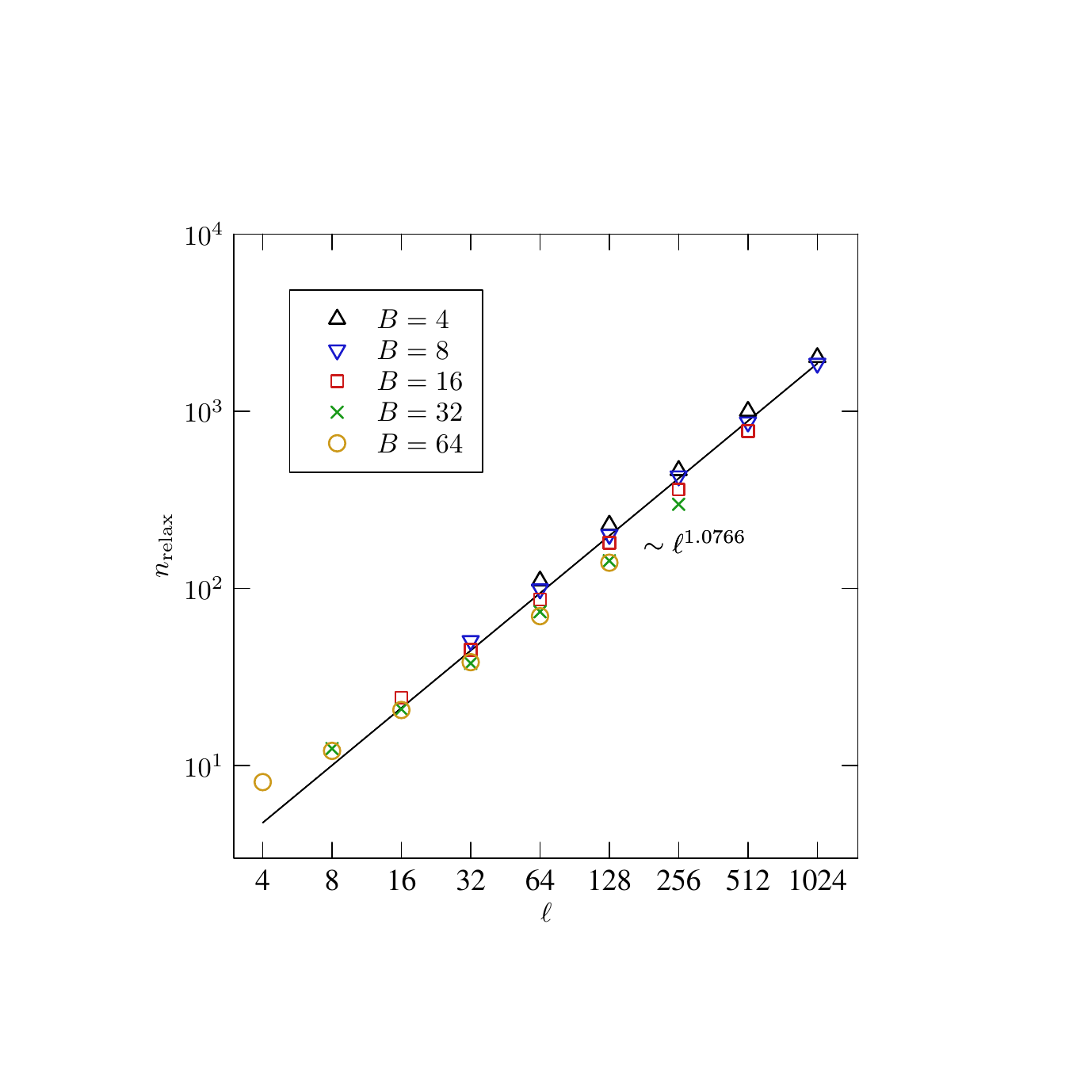}
  \caption{(Color online) Required number $n_\mathrm{relax}$ of iterations for the label relaxation
    technique for tile consolidation as a function of the renormalized system size
    $\ell = (L/B)$. The line is a fit of the form \eqref{eq:relax_scaling} to the data for
    $\ell \ge 100$, yielding $d_\mathrm{min} = 1.0766$.}
  \label{fig:relax_scaling}
\end{figure}

Cluster labels can be consolidated across tile boundaries using a relaxation
procedure similar to the self-labeling employed above inside of tiles
\cite{flanigan:95}. In a preparation step, for each edge crossing a tile boundary the
indices of the cluster roots of the two sites connected by the boundary-crossing bond
are stored in an array, cf.\ Fig.~\ref{fig:relax} (kernel {\tt prepare\_relax()}). In
the relaxation phase each tile sets the root labels of its own active boundary sites
to the minimum of its own current label and that of the corresponding neighboring
tile. Relaxation steps are repeated until local cluster labels do not change any
further. Similar to self-labeling, the number of relaxation steps scales as the
shortest path between two points on the largest cluster(s), however, the relevant
length scale for the relaxation procedure is now $\ell = L/B$, leading to the
following scaling behavior at the percolation threshold
\begin{equation}
  n_\mathrm{relax} \sim \ell^{d_\mathrm{min}}.
  \label{eq:relax_scaling}
\end{equation}
For systems below the transition temperature or more general cluster identification
tasks with extensive, but non-fractal clusters, $d_\mathrm{min}$ is replaced by 1,
whereas above the critical point and for other problems with finite characteristic
length scales $n_\mathrm{relax} \sim \xi/B$. The number of iterations
$n_\mathrm{relax}$ is shown for a simulation of the $q=2$ Potts model at criticality
in Fig.~\ref{fig:relax_scaling}. The expected scaling with $d_\mathrm{min} = 1.08$
\cite{miranda:91} is well observed for sufficiently large system sizes across all
tile sizes $B$: a fit of the functional form \eqref{eq:relax_scaling} results in
$d_\mathrm{min} = 1.0766$. The small, but visible downward shift of
$n_\mathrm{relax}$ with increasing $B$ results from concurrency effects: for a small
total number of tiles many of them are treated at the same time on different
multiprocessors, resulting in the possibility of a label propagating to tiles more
than one step away in one pass if (as is likely) several of the boundary sites belong
to the same clusters.

The number of operations per relaxation iteration is proportional to the length of
the tile boundary times the number of tiles, i.e.,
\begin{equation}
  t_\mathrm{relax} \sim B\ell^2.
\end{equation}
The relaxation routine (kernel {\tt relax()}) appears intrinsically serial in nature
as different boundary spins can point to the same roots such that concurrent
operations could lead to inconsistencies, unless appropriate locks are
used. Nevertheless, an alternative implementation (kernel {\tt relax\_multithread()})
using $B$ threads to update a number of boundary spin pairs concurrently in a thread
block is perfectly valid as similar to the self-labeling approach only the number of
necessary iterations is affected by the order of operations while the final result is
not changed.  As different blocks can essentially only be synchronized between kernel
calls, the stopping criterion is checked on CPU in between kernel invocations. The
parallel run time for this kernel is then given by
\begin{equation}
  \label{eq:relax_runtime}
  T^p_\mathrm{global} = C_\mathrm{relax} \frac{\ell^2}{\min(\ell^2, n)} \frac{B}{\min(B,m)}
  \ell^{d_\mathrm{min}}
\end{equation}
Note that the asymptotic effort per spin from the relaxation phase,
$T^p_\mathrm{global}/L^2 \sim B^{-1}\ell^{d_\mathrm{min}} \propto L^{d_\mathrm{min}}$,
does not become constant as the system size is increased, unless the tile size $B$ is
scaled proportionally to $L$.

For root finding in the spin-flipping phase, it is of some relevance that the
relaxation process effectively attaches all sub-clusters in tiles belonging to the
same global cluster directly to the root of the sub-cluster with the smallest cluster
label. Therefore, the algorithm involves path compression on the level of the coarse
grained lattice.

\subsubsection{Hierarchical sewing}

\begin{figure}[tb]
  \centering
  \includegraphics[width=0.7\columnwidth]{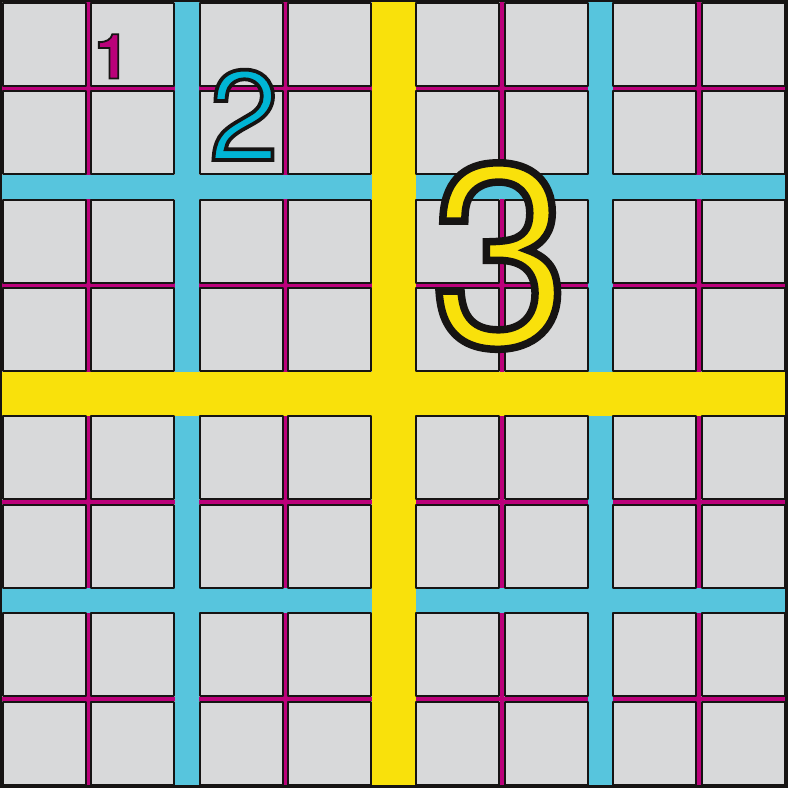}
  \caption{(Color online) Hierarchical sewing of $64$ tiles for label consolidation. On level 1,
    $2\times 2$ tiles are sewn together to form $16$ larger tiles. In levels 2 and 3,
    the tile numbers are reduced to $4$ and $1$, respectively.}
  \label{fig:hierarchical}
\end{figure}

An alternative technique of label consolidation across tiles uses a hierarchical or
divide-and-conquer approach as schematically depicted in Fig.~\ref{fig:hierarchical}
\cite{baillie:91}. On the first level $2\times 2$ tiles of $B\times B$ spins are sewn
together by inserting the missing bonds crossing tile boundaries. This results in
$B/2\times B/2$ larger tiles which are then combined in a second step etc.\ until,
finally, labels of the whole system have been consolidated. For the case of periodic
boundary conditions, the bonds wrapping around the lattice in both directions need to
be inserted in an additional step. Bond insertion itself is performed in the
union-and-find manner described above using tree balancing, i.e., the roots of the
two clusters connected by the added bond are identified and the smaller cluster is
then attached to the root of the larger cluster. We can assume that find times are
essentially constant inside of the original tiles of size $B$, either because tile
labeling was performed with the breadth-first or self-labeling algorithms which
produce labelings with complete path compression (i.e., each node label points
directly to the root), or since it was done using union-and-find with (at least) one
of the ingredients of tree balancing or path compression, leading to (at most)
logarithmic time complexity of finds. Then, using tree balancing in the hierarchical
sewing step ensures that find times remain logarithmically small as tiles are
combined. Time complexity could be further improved by adding path compression, but
(as for union-and-find inside of tiles) it is found here that this rather slows down
the code in the range of lattice sizes considered here. Note that the self-labeling
approach does not naturally provide the information about cluster sizes in the tree
roots. It is found, however, that it has no adverse effect on the performance of
the tile consolidation step if cluster sizes are simply assumed to be identical (and,
for simplicity, equal to one) for partial clusters inside of tiles.

\begin{figure}[tb]
  \centering
  \includegraphics[keepaspectratio=true,scale=0.8,trim=45 48 75 78]{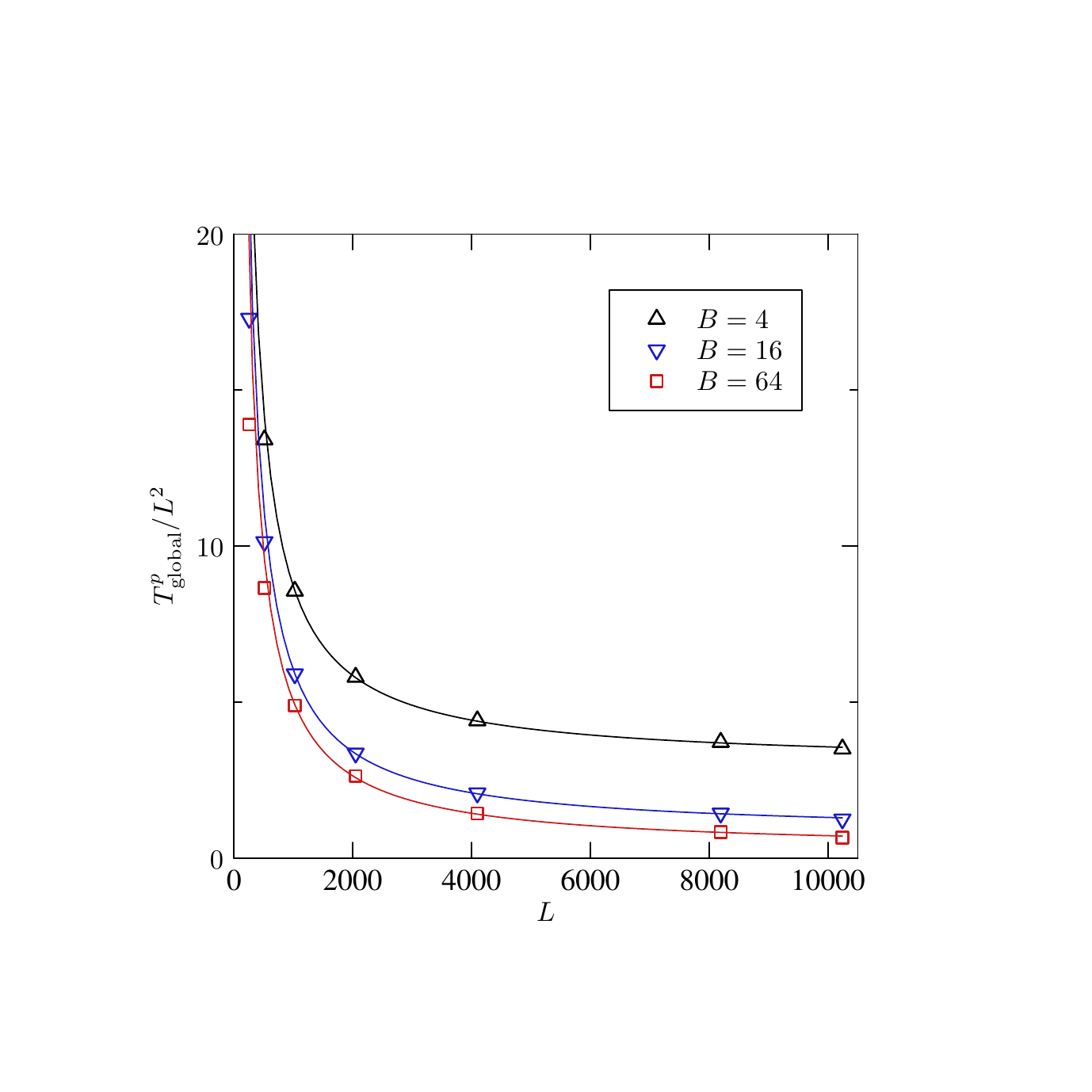}
  \caption{(Color online) Total parallel run times $T^p_\mathrm{global}$ for label consolidation
    via hierarchical sewing as a function of system size $L$ for different tile sizes
    $B$ for the $q=2$ critical Potts model on the GTX 480. The lines are fits of the
    form $T^p_\mathrm{global} = a/L+b/B$ to the data.  }
  \label{fig:hierarchical_scaling}
\end{figure}

One thread block is assigned to a configuration of $2\times 2$ tiles at each
level. The sewing itself is essentially serial in nature. For one of the two linear
seams of each sewing step (the horizontal seam, say), one can use two threads,
however, leading to two $2\times 1$ tiles after finishing the horizontal seam which
are only combined into one larger tile by closing the vertical seam\footnote{Note
  that this still leads to some under-utilization of the device due to the limit of
  eight active blocks per multiprocessor which requires at least four threads per
  block for full occupancy with blocks.}. As the tile size for the $k$th generation
is $B_k = 2^k B$ and the length of the seam is $2\times2^k B$, the serial
computational effort for level $n$ of the sewing is
\begin{equation}
  T^s_k = C_\mathrm{sew} \left(\frac{L}{2^k B}\right)^2 (2\times2^k B) =
  C_\mathrm{sew} \frac{L^2}{B} 2^{1-k}.
\end{equation}
where I have neglected logarithmic terms due to the find operations. The total number
of levels is $k_\mathrm{max} = \log_2(L/B)$ (assuming, for simplicity, that $L$
and $B$ are powers of two). Hence, the total serial effort is
\begin{equation}
  T^s_\mathrm{global} = \sum_{k=1}^{k_\mathrm{max}} T^s_k =
  \frac{L^2C_\mathrm{sew}}{B} \sum_{k=1}^{k_\mathrm{max}} \left(\frac{1}{2}\right)^{k-1}
  = \frac{2L^2 C_\mathrm{sew}}{B}\left(1-\frac{B}{L}\right),
\end{equation}
On the GPU device with $n$ multiprocessors mapped to independent blocks available for
the sewing procedure, the parallel run time for generation $k$ is
\begin{equation}
  T^p_k = \frac{T^s_k}{\min[(\ell 2^{-k})^2,n]}.
\end{equation}
For a sufficiently large system, at the beginning of the process the number of tiles
$(\ell 2^{-k})^2$ to sew will always exceed $n$. As the number of remaining tiles is
reduced, the number of sewing jobs will drop to reach the number of multiprocessors
at $(\ell 2^{-k^\ast})^2 = n$ or
\begin{equation}
  k^\ast = \log_2 \frac{\ell}{\sqrt{n}},
\end{equation}
where another approximation is made by allowing for non\hyp{}integer level numbers $k$.
Due to the variable number of multiprocessors actually involved in the calculation,
the total parallel effort has two contributions,
\begin{eqnarray}
  T^p_\mathrm{global} & = & \sum_{k=1}^{k^\ast} \frac{T^s_k}{n} +
  \sum_{k=k^\ast+1}^{k_\mathrm{max}} \frac{T^s_k}{(\ell 2^{-k})^2} \nonumber \\
  & = & C_\mathrm{sew} \frac{L^2}{n B}\frac{2^{k^\ast}-1}{2^{k^\ast}} +
  4 C_\mathrm{sew}B(2^{k_\mathrm{max}}-2^{n^\ast}) \nonumber\\
  & = & C_\mathrm{sew}L^2\left[\frac{1}{nB}+\left(4-\frac{5}{\sqrt{n}}\right)\frac{1}{L}\right].
  \label{eq:hierarchical_scaling}
\end{eqnarray}
Therefore, the effort $T^p_\mathrm{global}/L^2$ per site becomes asymptotically
independent of $L$, but this limit is approached rather slowly with a $1/L$
correction, whereas effects of incomplete loading of the device decay as $1/L^2$ (in
two dimensions). This is illustrated by the numerical results shown in
Fig.~\ref{fig:hierarchical_scaling}. The data are well described by the form
\begin{equation}
  \label{eq:hierarchical_fit}
  T^p_\mathrm{global}/L^2 = \frac{a}{L} + \frac{b}{B}
\end{equation}
expected from Eq.~\eqref{eq:hierarchical_scaling}. Comparing
Eqs.~\eqref{eq:hierarchical_scaling} and \eqref{eq:hierarchical_fit}, from the ratio
$a/b$ of fit parameters one can deduce the effective number $n$ of processing units
as
\begin{equation}
  n = \frac{25+8a/b+\sqrt{25+16a/b}}{32},
\end{equation}
and, for instance, the fit at constant $B=16$ yields $n\approx 110$, while a fit at
constant $L=8192$ results in $n\approx 113$, rather close to the theoretically
expected result for the GTX 480 with 8 blocks for each of the 15 multiprocessors,
resulting in $120$ processing elements. The somewhat smaller $n$ estimated are
attributed to effects of thread divergence and the neglect of logarithmic terms in
the find step. For tile size $B=16$, the asymptotic performance of this kernel if
found to be $T^p_\mathrm{global}/L^2 = 0.78$~ns.

\subsection{Cluster flipping}

\begin{figure}[tb]
  \centering
  \includegraphics[keepaspectratio=true,scale=0.8,trim=45 48 75 78]{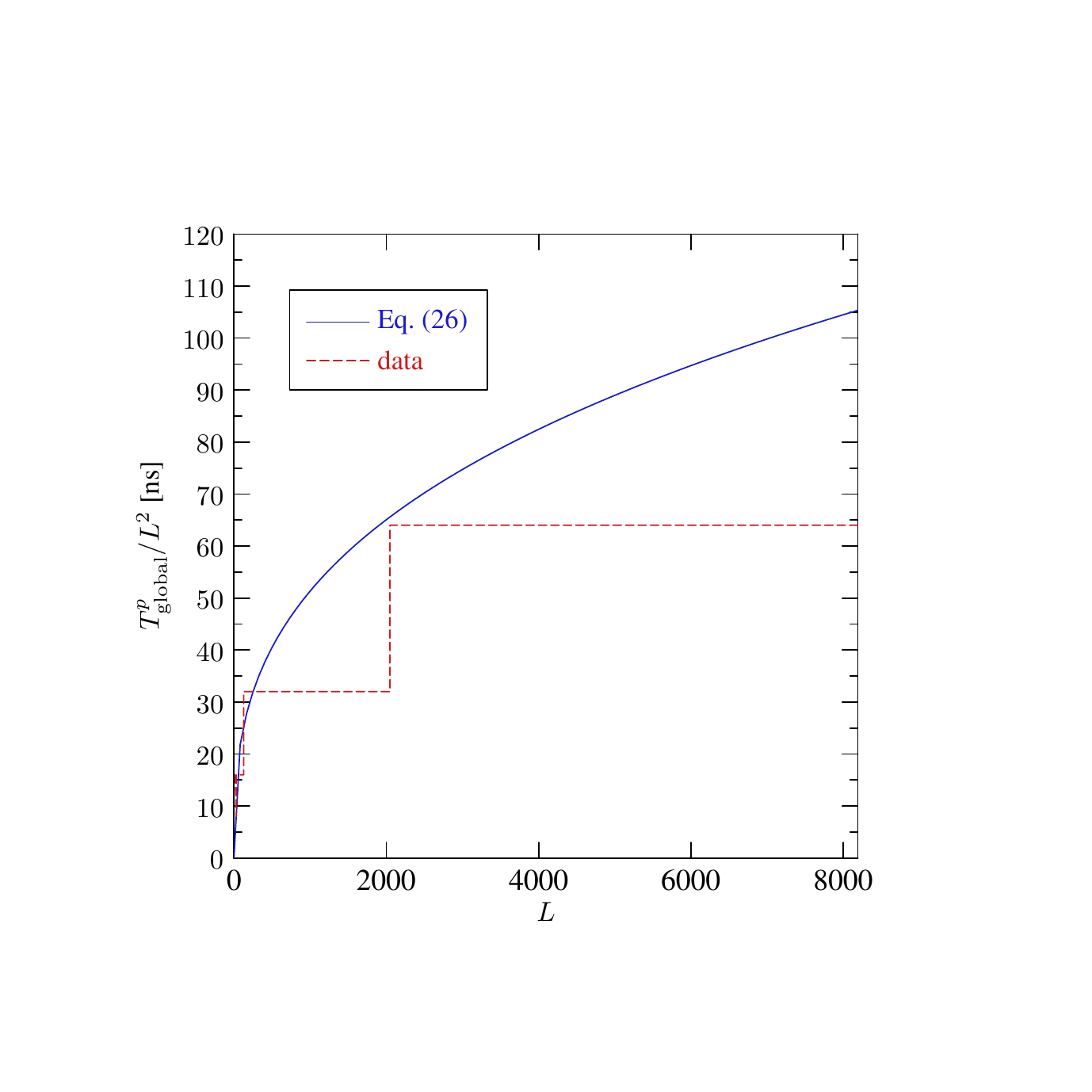}
  \caption{(Color online) Optimal tile size $B_\mathrm{opt}$ in cluster identification with
    self-labeling and label relaxation for the $q=2$ states critical Potts model as a
    function of $L$. The solid line shows the result of Eq.~\eqref{eq:optimaltile2}
    and the dashed line represents the optimum actually observed under the
    constraints $B\le 64$, $B=2^n$, $n=1$, $2$, $\ldots$.}
  \label{fig:relax_optimal_tiles}
\end{figure}

Having globally identified the connected components resulting from the bond
configuration, step 3 of the SW algorithm described at the beginning of
Sec.~\ref{sec:SW} consists of assigning new, random spin orientations to each cluster
and adapting the orientation of each spin in the cluster to the new orientation
prescribed. Since it is inconvenient to keep a separate list of global cluster roots,
it is easiest to generate a new random spin orientation for each lattice site while
only using this information at the cluster roots. To this end, the array of now
superfluous bond activation variables is re-used. In a first kernel, {\tt
  prepare\_flip()}, a random orientation is drawn and stored in the bond array for
each site. This is done in tiles of $B_x\times B_y$ sites as for the bond activation,
using the same array of random-number generators. In a second step (kernel {\tt
  flip()}), each site performs a find operation to identify its root and applies the
new spin orientation found there to the local spin. Since cluster labels are
effectively stored in a tree structure, this step involves non-local memory accesses
for each site. In principle, locality could be improved here by employing full path
compression in the union steps before, but in practice this is not found to improve
performance for the system sizes up to $16\,384\times 16\,384$ considered
here. Another possible improvement would eliminate the wasteful operation of drawing
new proposed orientations for all spins while only the new orientations of the
cluster roots are required. This can be achieved by carrying the flipping information
piggy-back on the cluster labels, at least for the $q=2$ or Ising model where
flipping information is only one bit wide. Again, however, in practice it is found
that due to the incurred complications in the arithmetics regarding cluster labels in
find and union operations, overall performance is actually decreased by this
``optimization''. Due to the necessary tree traversal, the performance of the cluster
flipping procedure depends weakly on the degree of path compression performed
previously in cluster labeling on tiles and label consolidation as well as on the
tile size $B$. For the combination of self-labeling on tiles and hierarchical sewing,
it is found to be $T^p_\mathrm{flip}/L^2 = 0.201$~ns for $L=8192$ and $B=16$, while
it is somewhat smaller at $0.133$~ns if label relaxation is used instead of
hierarchical sewing.

\subsection{Performance and benchmarks\label{sec:performance}}

\begin{figure}[tb]
  \centering
  \includegraphics[keepaspectratio=true,scale=0.8,trim=45 48 75 78]{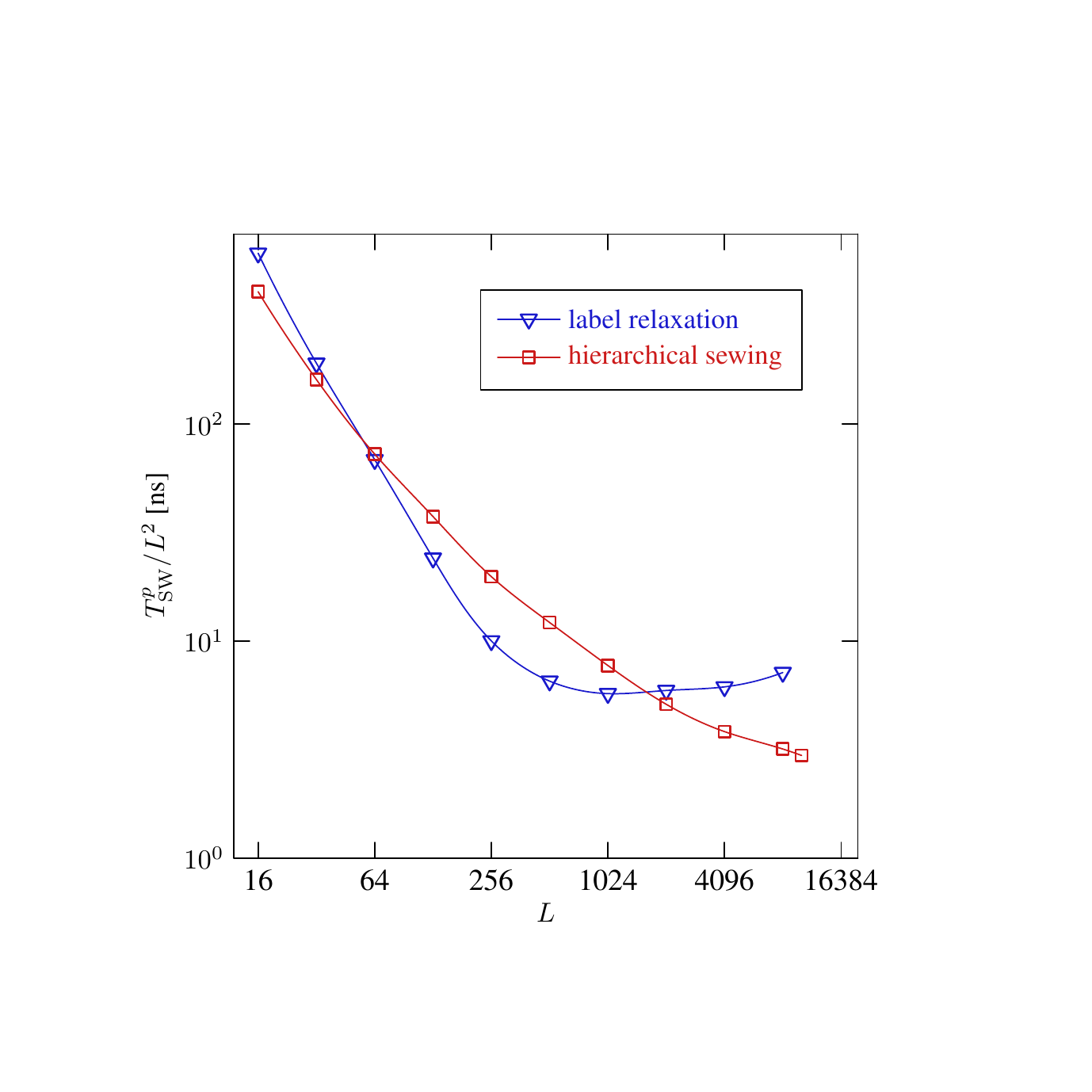}
  \caption{(Color online) Total run times $T^p_\mathrm{SW}$ per spin in nanoseconds for the
    Swendsen-Wang update of the $q=2$ critical Potts model on the GTX 480 from
    self-labeling on tiles plus label consolidation with label relaxation and
    hierarchical sewing, respectively. Lines are guides to the eye.}
  \label{fig:performance}
\end{figure}

As a number of options for the cluster identification task have been discussed, the
question arises which of them is the most efficient for a given set of parameters and
a given GPU device. For the bond activation and cluster flipping steps, the situation
is simpler as no important variants haven been discussed there, such that these steps
are always performed with the outlined local approaches and tiles with $B_x = 256$
and $B_y = 4$, apart from the smallest systems with $L < 256$. Regarding the cluster
labeling in tiles, it is clear from the data presented in
Fig.~\ref{fig:label_scaling} that self-labeling shows the best performance for block
sizes $B \lesssim 128$. The main decision is thus between the label relaxation and
hierarchical sewing approaches for label consolidation. Additionally, an optimal tile
size needs to be determined. For the combination of self-labeling and hierarchical
sewing, the total parallel run time for cluster identification is
\begin{equation}
  \label{eq:total_runtime}
  T^p_\mathrm{identify}/L^2 = \frac{C_\mathrm{local}}{mn} B^{d_\mathrm{min}}+
  \left(\frac{a}{L}+\frac{b}{B}\right),
\end{equation}
assuming that $B^2/k \ge m$ in Eq.~\eqref{eq:selflabel_scaling}. Here, $a$ and $b$
are the parameters from Eq.~\eqref{eq:hierarchical_fit}. On minimizing, the optimal
tile size is then found to be
\begin{equation}
  \label{eq:optimaltile1}
  B_\mathrm{opt} =
  \left(\frac{b}{d_\mathrm{min}C_\mathrm{local}/mn}\right)^{1/(d_\mathrm{min}+1)}.
\end{equation}
The fit parameters for the runs on the GTX 480 and $d_\mathrm{min} = 1.08$ then yield
$B_\mathrm{opt} \approx 14.2$. Since, for simplicity, runs were restricted to $L$ and
$B$ being powers of two, $B=16$ is closest to the optimum. Similar fits for the data
on the GTX 580 and the Tesla M2070 also used for test runs yield the same optimum in
the power-of-two step sizes. The pre-asymptotic branch with $B^2/k \le m$ in
Eq.~\eqref{eq:selflabel_scaling} does not yield an optimum, but total run times are
monotonously decreasing with $B$. In other words, as long as idle cores in the
multiprocessors are available, the tile size should be increased. $B=16$ hence is
also the global optimum for this setup.

\begin{figure}[tb]
  \centering
  \includegraphics[keepaspectratio=true,scale=0.8,trim=45 48 75 78]{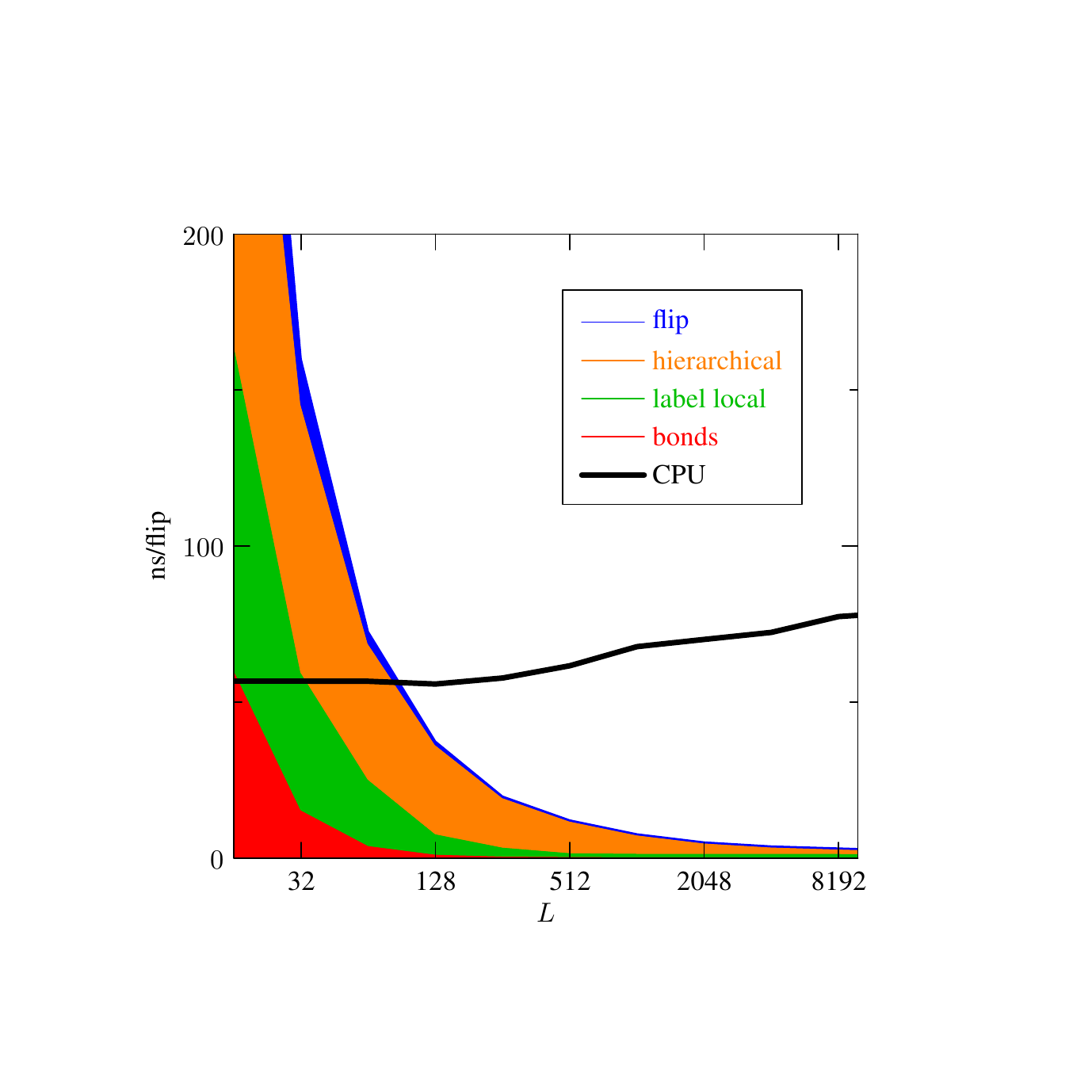}
  \caption{(Color online) Break down of parallel run times for one SW update per spin into the
    components of bond activation, local labeling, label consolidation via the sewing
    approach and spin flipping. The CPU curve shows reference data for a serial
    implementation running on an Intel Core 2 Quad Q6700 at 2.66 GHz.}
  \label{fig:cluster}
\end{figure}

For the combination of self-labeling and label relaxation, the total run time for
an update is
\begin{equation}
  \label{eq:total_runtime2}
  T^p_\mathrm{identify}/L^2 = \frac{C_\mathrm{local}}{mn} B^{d_\mathrm{min}}+
  \frac{C_\mathrm{relax}}{mn} \frac{L^{d_\mathrm{min}}}{B^{d_\mathrm{dim}+1}},
\end{equation}
such that the optimal tile size becomes
\begin{equation}
  \label{eq:optimaltile2}
  B_\mathrm{opt} =
  \left(\frac{C_\mathrm{relax}(d_\mathrm{min}+1)}{C_\mathrm{local}d_\mathrm{min}}
  L^{d_\mathrm{min}}\right)^{1/(2d_\mathrm{min}+1)},
\end{equation}
which (with $d_\mathrm{min}\approx 1.08$) is approximately proportional to $L^{1/3}$
for the critical $q=2$ Potts model in two dimensions. Figure
\ref{fig:relax_optimal_tiles} shows the resulting optimal tile size as a function of
$L$. Due to the limitation of shared memory to $48$ kB on current NVIDIA GPUs,
self-labeling on tiles is limited to block sizes $B\le 64$ (assuming $B=2^n$, $n=1$,
$2$, $\ldots$), such that the optimal tile sizes cannot be used for $L\gtrsim
4096$. Working directly in global memory is no option as it slows down the code
dramatically. Using breadth-first search or union-and-find on larger tiles is
feasible, but does significantly increase the total run time, even though the
relaxation phase is slightly more efficient. I therefore did not increase the tile
size beyond $B=64$, as indicated in Fig.~\ref{fig:relax_optimal_tiles}.

\begin{figure}[tb]
  \centering
  \includegraphics[keepaspectratio=true,scale=0.8,trim=45 48 75 78]{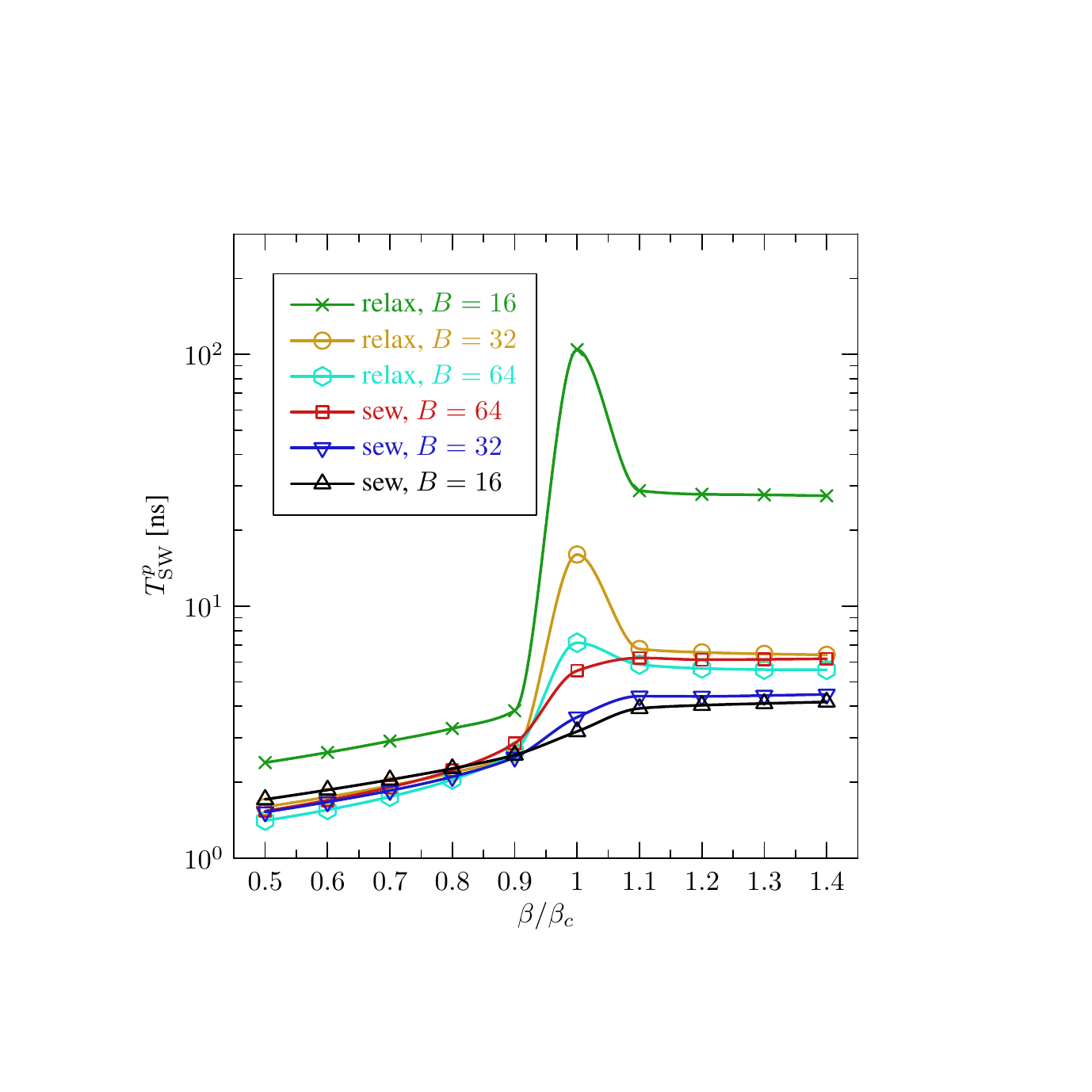}
  \caption{(Color online) Run times for the SW update on the GTX 480 as a function of the inverse
    temperature $\beta$ for the relaxation and sewing approaches and different tile
    sizes.}
  \label{fig:offcritical}
\end{figure}

The resulting total run times on the GTX 480 are shown in
Fig.~\ref{fig:performance}. The two consolidation approaches lead to quite different
size dependence. Tile relaxation results in a rather fast decay of run times per site
in the under-utilized regime and is faster than the sewing approach for intermediate
system sizes. Eventually, however, the scaling
$$
T^p_\mathrm{identify}/L^2  \sim L^\frac{d_\mathrm{min}^2}{2d_\mathrm{min}+1}
$$
implied by Eqs.~\eqref{eq:total_runtime2} and \eqref{eq:optimaltile2} kicks in, which
amounts to \linebreak $T^p_\mathrm{identify}/L^2 \sim L^{0.367}$ for
$d_\mathrm{min}=1.08$, and results in the upturn seen in
Fig.~\ref{fig:performance}. For the hierarchical approach, on the other hand, as
implied by Eq.~\eqref{eq:total_runtime} the best performance is reached only rather
slowly as $L$ is increased, but $T^p_\mathrm{identify}/L^2$ ultimately becomes
constant as (theoretically) $L\rightarrow\infty$. At $L=8192$ and $\beta = \beta_c$
for the $q=2$ Potts model, SW with sewing performs at $3.18$ ns per spin and per
sweep on the GTX 480, while relaxation results in $7.15$~ns per sweep. For the pure
cluster identification problem, i.e., without the bond activation and spin flipping
steps, these times are reduced to $2.52$~ns and $6.56$~ns, respectively. Figure
\ref{fig:cluster} shows the break down of run times into the algorithmic components
of bond activation, labeling on tiles, tile consolidation and spin flipping when
using hierarchical sewing. Label consolidation is the dominant contribution up to
intermediate system sizes, and only for $L \ge 16\,384$ its run time drops below that
of local labeling on tiles. For smaller systems, the fraction of time spent on bond
activation and spin flipping is negligible, while (due to the decrease in time spent
for label consolidation) it rises to about 20\% for $L=8192$. As a reference,
Fig.~\ref{fig:cluster} also shows the run time of an optimized, serial CPU
implementation using breadth-first search and on-line flipping of spins as the
clusters are grown, running on an Intel Core 2 Quad Q6700 at 2.66 GHz.

\begin{figure}[tb]
  \centering
  \includegraphics[keepaspectratio=true,scale=0.8,trim=45 48 75 78]{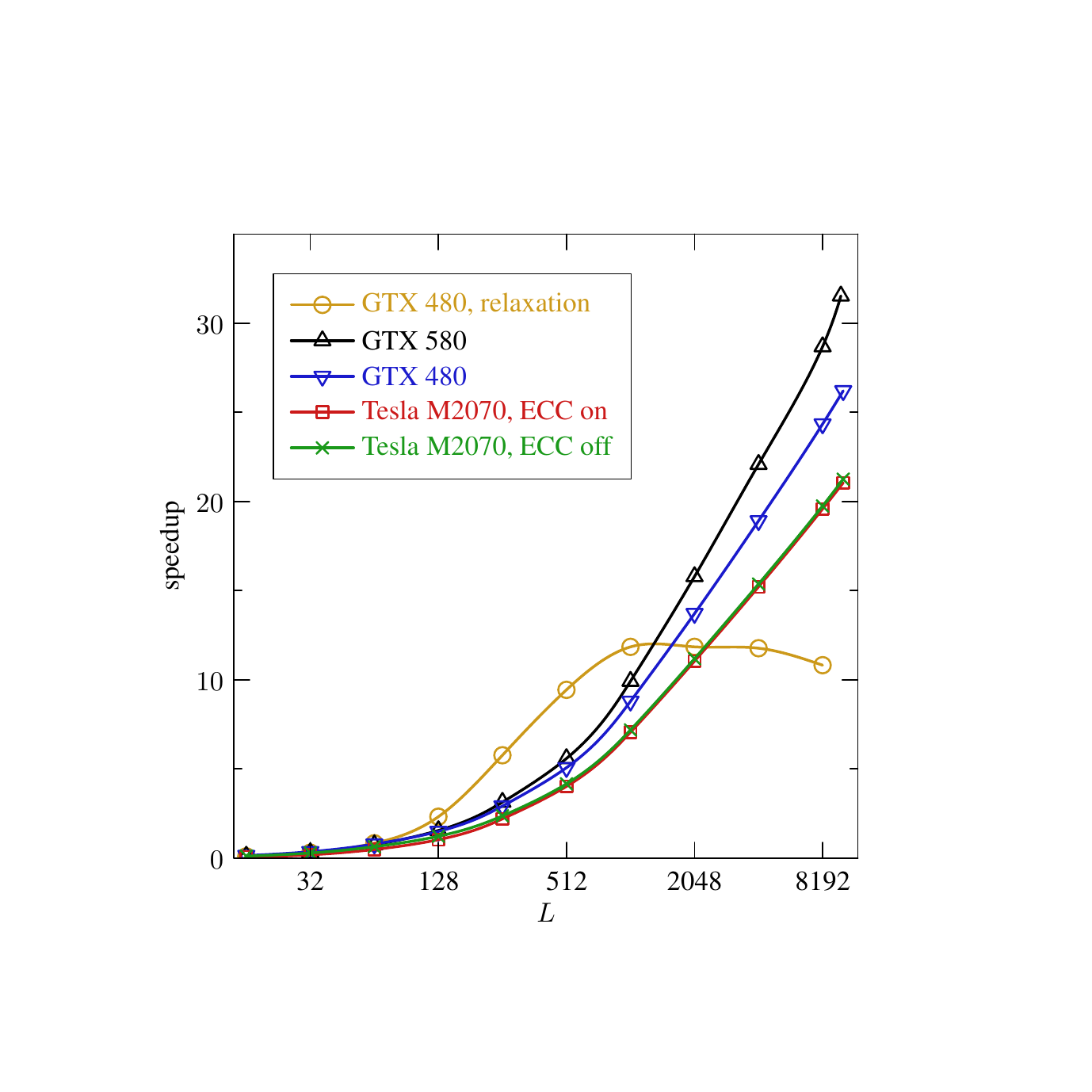}
  \caption{(Color online) Speed-up of the Swendsen-Wang update for the $q=2$ critical Potts model on
    GPU as compared to CPU (Intel Q6700 at 2.66 GHz) as a function of system size
    $L$. The circles show results from using the relaxation procedure while the
    remaining data sets are for the hierarchical sewing process on the GTX 480, GTX
    580 and Tesla M2070 GPUs, respectively. The latter is shown in variants with and
    without error-correcting code (ECC).}
  \label{fig:cluster_speedup}
\end{figure}

The incipient percolating clusters for the Potts model simulations at $\beta_c$ are
typical for a critical model. For other applications, for instance in image
segmentation, it is interesting to investigate the performance for more general
situations. Figure \ref{fig:offcritical} displays the run times for an SW update as a
function of inverse temperature $\beta$, comparing the setups with relaxation and
hierarchical sewing for tile consolidation. There is a natural increase in run time
with the concentration $p=1-\exp(-\beta J)$ of bonds. While for the sewing procedure,
run times increase monotonically with $\beta$, for the relaxation approach there is a
pronounced peak of run times near $\beta = \beta_c$, where the number of necessary
iterations $n_\mathrm{relax}$ shoots up since now information about the incipient
percolating cluster needs to be transmitted across the whole system. Run times become
somewhat smaller again for $\beta > \beta_c$ as most bonds crossing tile boundaries
belong to the same (percolating) cluster such that, due to concurrency, cluster
labels can travel several steps in one iteration. This concurrency effect strongly
increases as more tiles are treated simultaneously which, for a fixed number of
multiprocessors, is the case for larger tile sizes. Figure \ref{fig:offcritical}
shows that the preference of the sewing procedure over relaxation for large systems
is robust with respect to variations in temperature and should also be justified for
more general structures not resulting from a percolation transition.

\begin{table*}[tb]
  \caption{Benchmark results for the Swendsen-Wang update of the $q=2$, $q=3$ and $q=4$ Potts models 
    on GPU vs.\ CPU.}
  \label{tab:results}
  \centering
  \begin{tabular}{rr.ll.@{\,ns\hspace*{2em}}l.@{\,ns\hspace*{2em}}.} \hline
    $q$ & $\beta/\beta_c$ & $L$   & method                & \multicolumn{2}{c}{CPU} &
    \multicolumn{2}{c}{GPU} & \multicolumn{1}{c}{speedup}\\ \hline
    2   & 1               & $512$   & self-labeling, relaxation & Q6700 & 61.63 & GTX 480 & 6.533 & $9$ \\
    2   & 1               & $512$   & self-labeling, sewing & Q6700 & 61.63 & GTX 480 & 12.17 & $5$ \\
    2   & 1               & $8192$  & self-labeling, sewing & Q6700 & 77.39 & GTX 480 & 3.179 & $24$ \\
    2   & 1               & $8192$  & self-labeling, sewing & Q6700 & 77.39 & GTX 580 & 2.697 & $29$ \\
    2   & 1               & $8192$  & self-labeling, sewing & i7@9300 & 105.8 & GTX 580 & 2.697 & $39$ \\
    2   & 1               & $8192$  & self-labeling, sewing & Q6700 & 77.39 & M2070 & 3.934 & $20$ \\
    2   & 1               & $8192$  & self-labeling, sewing & E5620 & 149.6 & M2070 & 3.934 & $38$ \\
    2   & 1               & $16\,384$  & self-labeling, sewing & E5620 & 152.1 & M2070 & 3.573 & $43$ \\
    2   & 1               & $8192$  & self-labeling, relaxation & Q6700 & 77.39 & GTX 480 & 7.154 & $11$ \\
    2   & $0.6$           & $8192$  & self-labeling, sewing & Q6700 & 57.12 & GTX 480 & 1.863 & $31$ \\
    2   & $1.4$           & $8192$  & self-labeling, sewing & Q6700 & 135.7 & GTX 480 & 4.164 & $33$ \\
    3   & 1               & $8192$  & self-labeling, sewing & Q6700 & 70.73 & GTX 480 & 3.059 & $23$ \\
    4   & 1               & $8192$  & self-labeling, sewing & Q6700 & 65.51 & GTX 480 & 2.887 & $23$ \\ \hline
  \end{tabular}
\end{table*}

Figure \ref{fig:cluster_speedup} shows the speed-up of the GPU implementation with
respect to the CPU code on the Q6700 processor. For large systems, speed-ups in
excess of 30 are observed. Comparing different GPU devices, a clear scaling with the
number of multiprocessors and global memory bandwidth is observed with the best
performance seen for the GTX 580 ($n=16$, 192 GB/s), followed by the GTX 480 ($n=15$,
177 GB/s) and the Tesla M2070 ($n=14$, 144 GB/s). Naturally, effects of higher
double-precision floating-point performance of the latter are not seen for the
present code, which almost exclusively uses integer and a few single-precision
floating point arithmetic instructions. The penalty for activating error correction
(ECC) on the memory is minute. Some benchmark results, also including different
processors, are collected in Tab.~\ref{tab:results}.

%\begin{itemize}
%\item give performance for cluster identification separately
%\item memory boundedness of cluster identification (random access)
%\end{itemize}

\section{Wolff algorithm\label{sec:single}}

For simulations of spin models, Wolff \cite{wolff:89a} suggested a variant of the
Swendsen-Wang algorithm where only a single cluster, seeded at a randomly chosen
site, is grown at a time which is then always flipped. Empirically, it is found that
this leads to somewhat smaller autocorrelation times than SW
\cite{wolff:89,baillie:91a} but, most likely, no change in the dynamical critical
exponent (at least for integer $q$) \cite{deng:09}. Conceptually, one can imagine the
single-cluster algorithm as a variant of the SW dynamics where after a full
decomposition of the lattice according to the SW prescription, a site is picked at
random and the cluster of spins it belongs to is flipped. Since the probability of
picking a specific cluster in this way is proportional to its size, in this approach
larger clusters are flipped on average than in the original SW algorithm. This
explains the somewhat reduced autocorrelation times.

While this approach is easily coded in a serial program and, in addition to the
smaller autocorrelation times, in a suitable implementation performs at even somewhat
less effort per spin than the SW algorithm, it is not straightforwardly parallelized
\cite{evertz:93,bae:95,bader:06,kaupuzs:10}. The only obvious parallelism lies in the
sites at the wave front of the growing cluster, cf.\ the sketch in
Fig.~\ref{fig:BFS}. A number of approaches for parallel calculations come to mind:
\begin{itemize}
\item[(a)] A full parallel cluster labeling as in SW, followed by picking out and
  flipping a single cluster. Although many operations are wasteful here, there might
  still be a speed-up as compared to the serial code. If using a relaxation procedure
  for label consolidation, this approach can be somewhat improved by modifying the
  stopping criterion to only focus on the labels belonging to the cluster to be
  flipped.
\item[(b)] Restriction to wave-front parallelism \cite{evertz:93}. Due to the rather
  variable number of sites at the front, however, this generically leads to poor load
  balancing between the processing units. Load balancing can be improved by a
  de-localization of the wave front with a ``randomized'' rearrangement of the
  lattice. This can be reached, for instance, with a scattered strip partitioning,
  where strips of the lattice are assigned to available processing units in a
  round-robin fashion, leading to a more even distribution of sites at the wave front
  to different processors \cite{bae:95}.
\item[(c)] Suitable modifications of the single-cluster algorithm to make it more
  appropriate for parallel computation.
\end{itemize}
The approach (a) can be easily realized with the techniques outlined in
Sec.~\ref{sec:SW}. As discussed in Ref.~\cite{bae:95}, additional load balancing can
result in significant improvements on MIMD (multiple instruction, multiple data)
machines. It appears less suitable for the mixed architecture of GPUs. In contrast to
the more general case of SW dynamics discussed above in Sec.~\ref{sec:SW}, I refrain
here from a comprehensive evaluation of options, and only give some preliminary
results for a modification (c) of the Wolff algorithm appearing suitable for GPU
computing.

In this approach, the lattice is again decomposed into tiles of edge length $B$. A
single cluster {\em per tile\/} is then grown using a number of threads per tile to
operate on the wave front. Unlike the case of the SW implementation, the clusters are
{\em not\/} confined to the tiles, but can grow to span the whole lattice. One can
easily convince oneself, that the underlying decomposition remains to be the SW
one. If seeds in different tiles turn out to belong to the same cluster, different
parts of that cluster receive different labels, but since all clusters are flipped
the effect is the same as if a single cluster (for that two seeds) had been grown
(this is for the case of the $q=2$ model). Logically, this algorithm is identical to
performing the full SW decomposition and then selecting $\ell^2$ points on the
lattice, followed by flipping all distinct clusters these points belong to. While
this approach satisfies detailed balance (the SW decomposition remains the same and
the cluster flipping probability is symmetric), it is not ergodic as it stands since,
for instance, it becomes impossible to flip only a single spin. This deficiency can
be easily repaired, however, by assigning a flipping probability $p_\mathrm{flip} <
1$ to the clusters which can be large, but must be strictly smaller than one. If only
a relatively small number of tiles is chosen, the decorrelation efficiency of this
``few-cluster'' variant of the SW algorithm is about the same as that of the
single-cluster variant.

\begin{figure}[tb]
  \centering
  \includegraphics[keepaspectratio=true,scale=0.8,trim=45 48 75 78]{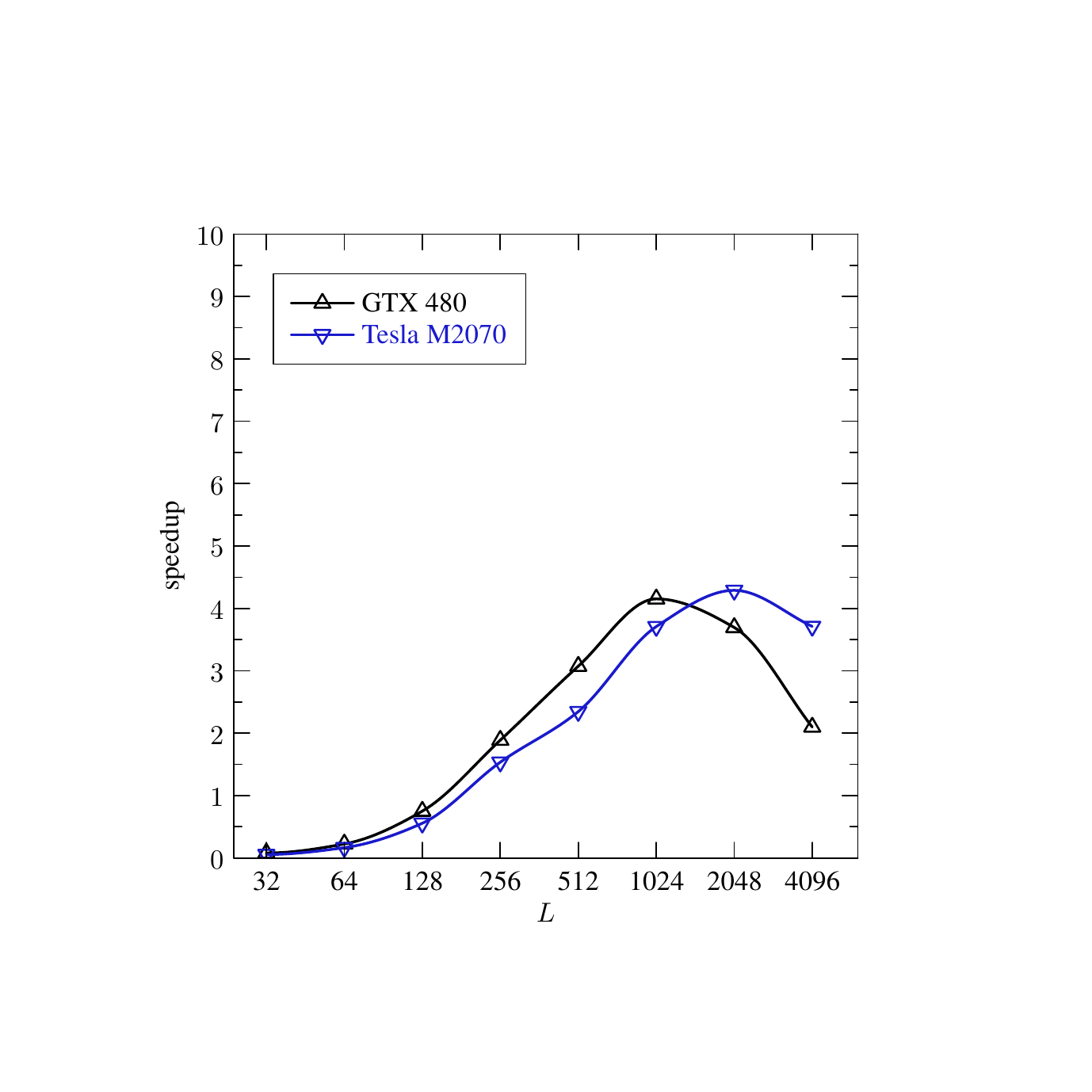}
  \caption{(Color online) Speed-up of the ``few-cluster'' update described in Sec.~\ref{sec:single}
    implemented on GPU as compared to a single-cluster update on CPU. For each system
    size, the optimal tile size $B$ has been selected from the range of allowable
    tile sizes determined by memory constraints.}
  \label{fig:single_speedup}
\end{figure}

For implementing the labeling in tiles, a number of threads $p$ per block is
chosen. If there are enough pending sites in the queue, each thread is assigned one
of these spins which are then examined in parallel. The queue is here realized as a
simple linear array of size $N=L^2$. This appears inefficient as the size of the wave
front will at most be of the order of $L^{d_H}$, where $d_H$ is the fractal dimension
of the cluster boundary. In contrast to the use of a ring buffer of length $\propto
L^{d_H}$, however, storing in and retrieving from the queue can be realized with
atomic operations only \cite{cuda}, i.e., without the use of locks. Unfortunately,
this setup severely limits the range of realizable tile sizes for larger systems as
memory requirements for this queue scale as $\ell^2 N = L^4B^{-2}$. In contrast to
the SW algorithm, bond activation and spin flipping can be done online with the
labeling pass. Consequently, the ``few-cluster'' implementation needs only two
kernels, {\tt cluster\_tile()} for the labeling and flipping and {\tt
  reset\_inclus()} for resetting the cluster labels after each pass. The number $p$
of threads per block is adapted to maximize occupancy of the device. In general, it
is found that good results are obtained on setting
\begin{equation}
  \label{eq:occupancy}
  p = \min\left[1024,1536/\min\left(\max(\ell^2/n,1),8\right)\right],
\end{equation}
as $1024$ is the maximum number of threads per block, $1536$ is the maximum number of
active threads and $8$ the maximum number of resident blocks per
multiprocessor. (Here, $n$ denotes the number of multiprocessors of the device.) The
resulting speed-ups as compared to a serial code on the Intel Core 2 Quad Q6700 are
shown in Fig.~\ref{fig:single_speedup}. The performance for large system sizes is
limited by the memory consumption of the queues, limiting the number $\ell^2$ of
tiles. Speed-ups by a factor of up to about five are achieved, significantly lower
than for the SW dynamics. It is expected than further optimizations (such as the use
of ring buffers instead of queues) could approximately double this
speed-up. Nevertheless, for cluster-update simulations on GPUs it might be more
efficient to stick with the SW algorithm.

\section*{Conclusions}

Cluster identification is a pivotal application in scientific computations with
applications in the simulation of spin models and percolation, image processing or
network analysis. While the underlying problem is inherently non-local in nature, the
choice of appropriate algorithms for implementations on GPU allows for significant
performance gains as compared to serial codes on CPU. The overall speed-up is seen to
be lowest for spin models at criticality, where clusters are fractal and span the
system. In all cases, however, speed-ups up to about 30 can be achieved on current
GPU devices. This is to be contrasted to the case of purely local algorithms, such as
Metropolis simulations of spin models, where speed-ups are seen to be larger by a
factor three to five \cite{weigel:10c,weigel:10a,weigel:11}. Even with this caveat,
it seems clear that GPU computing is not limited to the case of purely local problems
as significant performance gains can be achieved for highly non-local problems
also. Generalizations within the realm of spin-model simulations, such as variants on
different lattices or embedded clusters for O($n$) spin models \cite{wolff:89a} are
straightforward.

While the considerations presented here have been restricted to calculations on a
single GPU, it should be clear that the approach outlined for the Swendsen-Wang
dynamics or the pure cluster identification problem is easily parallelized across
several GPUs. For the case of spin-model simulations, the combination of
self-labeling and label relaxation appears better suited for this task as for the
final spin-flipping step only information local to each GPU is required, whereas for
the hierarchical scheme cluster roots (and therefore spin-flipping states) are
scattered throughout the whole system. The most effective setup for simulating large
systems, therefore appears to be the combination of self-labeling and hierarchical
sewing inside of a GPU and label relaxation between GPUs which can easily be realized
using MPI with rather low communication overheads.

\section*{Acknowledgments}

%I am indebted to T.~Yavors'kii for a careful reading of the manuscript.
%
Support by the DFG through the Emmy Noether Programme under contract No.\ WE4425/1-1
and by the Schwerpunkt f\"ur rechnergest\"utzte Forschungsmethoden in den
Naturwissenschaften (SRFN Mainz) is gratefully acknowledged.


%\bibliography{citeulike_nourl_noissn}

%Merlin.mbs v4.21 2009-07-09.
%


\end{document}